\documentclass[3p,12pt]{elsarticle}

\usepackage{color}
\usepackage{amsmath}
\usepackage{bm}
\usepackage{url}
\usepackage{scalerel,stackengine,wasysym} 
\usepackage{siunitx}

\newcommand{\torange}[1]{#1}

\newcommand{\tred}[1]{#1}

\newcommand{\tblue}[1]{#1}

\newcommand{\tgreen}[1]{#1}

\newcommand{\divB}{{\nabla}{\cdot}\bm{B}}
\newcommand{\plotbig}[1]{\includegraphics[width=1.0\textwidth]{#1}}
\newcommand{\plotone}[1]{\includegraphics[width=0.8\textwidth]{#1}}
\newcommand{\plotsml}[1]{\includegraphics[width=0.55\textwidth]{#1}}

\newcommand\vwtilde[2]{\ThisStyle{
  \setbox0=\hbox{$\SavedStyle#1$}
  \stackengine{-.1\LMpt}{$\SavedStyle#1$}{
    \stretchto{\scaleto{\SavedStyle\mkern.2mu\AC}{.5150\wd0}}{.6\ht0}
  }{O}{c}{F}{T}{S}^{#2}
}}

\journal{Journal of Computational Physics}

\begin{document}

\begin{frontmatter}

\title{
  Energy-consistent finite difference schemes \\
  for compressible hydrodynamics and magnetohydrodynamics \\
  using nonlinear filtering
}

\author{
  Haruhisa Iijima
}
\ead{h.iijima@isee.nagoya-u.ac.jp}
\address{
  Institute for Space-Earth Environmental Research, Nagoya University, Furocho, Chikusa-ku, Nagoya, Aichi 464-8601, Japan
}
\address{
  National Astronomical Observatory of Japan, 2-21-1 Osawa, Mitaka, Tokyo 181-8588, Japan
}

\begin{abstract}
  \torange{In this paper}, an energy-consistent finite difference scheme for the compressible hydrodynamic and magnetohydrodynamic (MHD) equations \tred{is introduced}.
  For the compressible magnetohydrodynamics, an energy-consistent finite difference formulation is derived using the product rule for the spatial difference.
  The conservation properties of the internal, kinetic, and magnetic energy equations can be satisfied in the discrete level without \torange{explicitly} solving the total energy equation.
  The shock waves and discontinuities in the numerical solution are stabilized by nonlinear filtering schemes.
  An energy-consistent discretization of the filtering schemes is also derived by introducing the viscous and resistive heating rates.
  The resulting energy-consistent formulation can be implemented with the various kinds of central difference, nonlinear filtering, and time integration schemes.
  \torange{
    The second- and fifth-order schemes are implemented based on the proposed formulation.
  }
  The conservation properties and the robustness of the present schemes are demonstrated \torange{via} one- and two-dimensional numerical tests.
  The proposed schemes successfully handle the most stringent problems in extremely high \tred{Mach number} and \tred{low beta} conditions.
\end{abstract}

\begin{keyword}
  Hydrodynamics
  \sep
  Magnetohydrodynamics
  \sep
  Computational plasma physics
  \sep
  Finite differences
  \sep
  Shock capturing
  \sep
  Spatial filtering
  \sep
  Skew-symmetric form
  \sep
  Secondary conservative
\end{keyword}

\end{frontmatter}



\section{Introduction}
\label{sec:intro}

The fully conservative finite difference scheme has been used as a tool for the numerical simulation of the incompressible flow \torange{due to its} very small numerical dissipations and the stability in the long-term simulations \cite{1965PhFl....8.2182H,1998JCoPh.143...90M,2004JCoPh.197..686M}.
In such schemes, the quadratic split form (also known as the skew-symmetric split form) of the convective term is used to achieve the kinetic energy conservation in the discrete sense while maintaining the momentum conservation without directly solving the kinetic energy equation.
The quadratic and cubic split operators can also reduce the aliasing error produced in the product of the multiple variables \cite{1997JCoPh.131..310K,2008JCoPh.227.1676K}.

\torange{Several researchers have attempted to extend} the split form of the convective term to the \torange{low Mach number} flows (see a review by Pirozzoli \cite{2011AnRFM..43..163P}).
The Jameson-Schmidt-Turkel scheme \cite{1981fpd..conf.....J} implicitly used the quadratic split operator in the finite volume formulation \cite{2000JCoPh.161..114D}.
The kinetic energy preserving (KEP) scheme is constructed using the quadratic split operator \cite{jameson2008construction,jameson2008formulation,2010JCoPh.229.7180P}.
The quadratic split operator can even achieve the consistency between the internal \torange{energy} and kinetic energy in the discrete sense \cite{2009JCoPh.228.6811K,2010JCoPh.229..276M}.
The concept of the quadratic split operator can be extended into the \torange{nonuniform} cylindrical coordinates \cite{2008JCoPh.227.7125D}.
The temporal derivatives can be also discretized using the quadratic split operator to \torange{obtain} the spatiotemporal conservation properties \cite{2010JCoPh.229..276M}.
The conservation of the entropy can also be achieved in addition to the energy-consistency \cite{2018JCoPh.375..823K}.

Fewer studies investigated the application of the energy-consistent schemes to the \torange{high Mach number} flows.
In \torange{high-speed flow simulations}, additional numerical diffusion schemes are required to stabilize the shock waves and discontinuities.
Jameson et al. \cite{1981fpd..conf.....J} used an artificial viscosity to stabilize their skew-symmetric finite volume scheme.
Ducros et al. \cite{2000JCoPh.161..114D} suggested the hybridization between the central and upwind schemes.
\torange{
Yee \cite{1999JCoPh.150..199Y} employed a nonlinear filtering approach for the entropy conserving scheme.
}
However, the energy consistency of these diffusion schemes \torange{was} rarely discussed explicitly, except \torange{in} the work by Shiroto et al. \cite{2018JCoPh.364....1S}.

In recent years, the application of the quadratic split operator has \torange{increasingly} been applied to the plasma physics, to achieve the secondary conservation properties.
Ni et al. \cite{2007JCoPh.227..174N,2007JCoPh.227..205N,2012JCoPh.231..281N} constructed the \torange{current density} conservative scheme for the incompressible magnetohydrodynamic (MHD) simulations.
Shiroto et al. \cite{2018JCoPh.364....1S} applied the approach to the two-temperature plasma flows.
There are several applications of the quadratic split operator on the kinetic plasma equations \cite{2007JCoPh.226..244I,2019JCoPh.379...32S}.
However, an energy-consistent scheme for application to compressible MHD simulations has not yet been constructed.

In the compressible MHD simulations, the consistency between the internal, kinetic, and magnetic energies is essential \torange{to accurately reproduce} the plasma dynamics and energetics.
\torange{Many numerical schemes solve the total (internal + kinetic + magnetic) energy equation} to satisfy the jump condition near the shocks and discontinuities.
However, in such schemes, the discretization errors of the kinetic and magnetic energy equations are imposed on the internal energy.
As a result, the evolution of the internal energy becomes inevitably erroneous when the internal energy is much smaller than the kinetic or magnetic \torange{energy}.
Such numerical simulations yield \tgreen{energetically inaccurate} solutions, such as the negative pressure.

\torange{This} study \torange{aims to construct} an energy-consistent \torange{formulation} for the compressible hydrodynamic and MHD equations \torange{while taking account the nonlinear filtering}.
Hence, an fully energy-consistent finite difference scheme is devised.
Our implementation of this scheme with the appropriate nonlinear filtering scheme provides superior numerical robustness, especially in the extremely high Mach number and low plasma beta (internal over magnetic energies) conditions.

This paper is organized as follows.
\torange{Section \ref{sec:scheme} presents the} energy-consistent formulations of the governing equations \torange{as well as} the nonlinear filtering flux.
Note that the proposed formulation is independent from the details of the finite difference operators, the filtering flux, and the temporal integration method.
\torange{Section \ref{sec:implement} describes} our implementation of the proposed scheme.
\tblue{
In Section \ref{sec:test}, several numerical tests are performed, especially focusing on the energy-consistency and numerical robustness of the proposed scheme.
}
Finally, Section \ref{sec:cncl} presents the conclusions of this article.

\section{Energy-consistent scheme for compressible hydrodynamic and  MHD equations}
\label{sec:scheme}

\subsection{Basic equations and analytical requirements}

\tblue{The basic equations of the ideal, fully compressible MHD system} in the Cartesian coordinates $x_i$ ($i=1,2,3$) can be written in a conservative form as
\begin{equation}
  \label{eq:anl_cont}
  \frac{\partial{\rho}}{\partial{t}}
  +\frac{\partial{\rho}{V_j}}{\partial{x_j}}
  =0,
\end{equation}
\begin{equation}
  \label{eq:anl_mom}
  \frac{\partial{\rho}{V_i}}{\partial{t}}
  +\frac{\partial{\rho}{V_j}{V_i}}{\partial{x_j}}
  +\frac{\partial{P}}{\partial{x_i}}
  +\frac{1}{2}\frac{\partial{B_jB_j}}{\partial{x_i}}
  -\frac{\partial{B_iB_j}}{\partial{x_j}}
  =0
  ,
\end{equation}
\begin{equation}
  \label{eq:anl_etot}
  \frac{\partial}{\partial{t}}
  \left(e+\frac{1}{2}{\rho}V_iV_i+\frac{1}{2}B_iB_i\right)
  +\frac{\partial}{\partial{x_j}}
  \left[
    \left(
      e+P+\frac{1}{2}{\rho}V_iV_i+B_iB_i
    \right){V_j}
    -B_jB_iV_i
  \right]
  =0,
\end{equation}
\begin{equation}
  \label{eq:anl_induc}
  \frac{\partial{B_i}}{\partial{t}}
  +\frac{\partial\left({V_j}{B_i}-{V_i}{B_j}\right)}{\partial{x_j}}
  =0.
\end{equation}
Here, the summation rule is assumed.
\tblue{The variables} $\rho$, $e$, and $P$ represent the mass density, the internal energy density, and the gas pressure, respectively, and $V_i$ and $B_i$ represent the $x_i$-component of the velocity field and the magnetic field, respectively.
The absence of the magnetic monopole provides a constraint on the magnetic field:
\begin{equation}
  \label{eq:anl_divb}
  \frac{\partial{B_j}}{\partial{x_j}}=0.
\end{equation}
The above equations are closed by the equation of states for the gas pressure $P(\rho,e)$.
The hydrodynamic equations can be derived by neglecting all components of the magnetic field (setting $B_i=0$ for any $i$).

The conservation law of the momentum expressed in Eqs. (\ref{eq:anl_mom}) can be rewritten as
\begin{equation}
  \label{eq:anl_mom_f}
  \frac{\partial{\rho}{V_i}}{\partial{t}}
  +\frac{\partial{\rho}{V_j}{V_i}}{\partial{x_j}}
  =
  F^P_i+F^L_i
  ,
\end{equation}
where $F^P_i$ is the pressure gradient force in the $i$-th direction
\begin{equation}
  F^P_i=-\frac{\partial{P}}{\partial{x_i}}
\end{equation}
and $F^L_i$ is the Lorentz force in the $i$-th direction
\begin{equation}
  \label{eq:anl_fl_cnsv}
  F^{L,c}_i=
  -B_j\left(
    \frac{\partial{B_j}}{\partial{x_i}}
    -\frac{\partial{B_i}}{\partial{x_j}}
  \right)
  +B_i\frac{\partial{B_j}}{\partial{x_j}},
\end{equation}
respectively.
The last term of Eq. (\ref{eq:anl_fl_cnsv}) \torange{disappears} when the solenoidal condition of Eq. (\ref{eq:anl_divb}) is satisfied.
The resultant Lorentz force can be written as
\begin{equation}
  \label{eq:anl_fl_ncnsv}
  F^L_i=
  -B_j\left(
    \frac{\partial{B_j}}{\partial{x_i}}
    -\frac{\partial{B_i}}{\partial{x_j}}
  \right).
\end{equation}
\torange{In this form,} the Lorentz force is normal to the direction of the magnetic field ($F^L_iB_i=0$).

In numerical simulations, the solenoidal constraint of the magnetic field is sometimes violated \torange{due to} discretization error.
Violation of \torange{the} $\divB=0$ condition produces the inconsistency between the conservative and non-conservative Lorentz force, which yields artificial field-aligned forces $B_i\divB$.

From \torange{the basic equations (Eqs. (\ref{eq:anl_cont})-(\ref{eq:anl_induc}))}, three independent energy equations, namely, internal, kinetic, and magnetic energy equations, \torange{can be derived:}
\begin{equation}
  \label{eq:anl_eint}
  \frac{\partial{e}}{\partial{t}}
  +\frac{\partial\left(e+P\right){V_j}}{\partial{x_j}}
  =-W^P,
\end{equation}
\begin{equation}
  \label{eq:anl_ekin}
  \frac{\partial\left({\rho}V_iV_i/2\right)}{\partial{t}}
  +\frac{\partial\left({\rho}V_iV_iV_j/2\right)}{\partial{x_j}}
  =W^P+W^L,
\end{equation}
\begin{equation}
  \label{eq:anl_emag}
  \frac{\partial\left(B_iB_i/2\right)}{\partial{t}}
  +\frac{\partial\left(B_iB_iV_j-B_jB_iV_i\right)}{\partial{x_j}}
  =-W^L,
\end{equation}
respectively.
Here, we defined the work done by the pressure gradient force
\begin{equation}
  W^P=V_iF^P_i
\end{equation}
and the work done by the Lorentz force
\begin{equation}
  W^L=V_iF^L_i.
\end{equation}
\torange{It is clear that} the equations for the internal, kinetic, and magnetic energies (Eqs. (\ref{eq:anl_eint})--(\ref{eq:anl_emag})) \torange{consist} of terms pertaining to the energy transport (i.e., the divergence of the enthalpy, kinetic energy, and Poynting \torange{fluxes}) and the \torange{works} done by the pressure gradient and Lorentz forces ($W^P_i$ and $W^L_i$).
Summation of \torange{the} three energy equations retrieves the law of the total energy conservation expressed in Eq. (\ref{eq:anl_etot}).

\torange{The momentum and magnetic flux conservation laws are solved in} \tblue{many} numerical schemes that employ the divergence formulation of the MHD equations.
However, the internal, kinetic, and magnetic energy equations are not solved as independent equations and, thus, the consistency among them is sometimes violated.
In stringent problems, violation of the energy-consistency sometimes yields a negative pressure.

\subsection{Discrete operators}

This work essentially follows the notation for the finite difference operators given in Morinishi et al. \cite{1998JCoPh.143...90M}.
The finite difference operator with stencil $n$ acting on a variable \tblue{$\Phi$} with respect to $x_1$ is defined by
\tblue{
\begin{equation}
  \label{eq:op_diff_n}
  \frac{{\delta_{n}}{\Phi}}{{\delta_{n}}{x_1}}
  \equiv
  \frac{\Phi\left(x_1+nh_1/2,x_2,x_3\right)
  -\Phi\left(x_1-nh_1/2,x_2,x_3\right)}
  {nh_1},
\end{equation}
}
where $h_1$ is the grid spacing in the $x_1$-direction.
The interpolation operator with stencil $n$ with respect to $x_1$ is defined as
\tblue{
\begin{equation}
  \label{eq:op_interp_n}
  \overline{\Phi}^{nx_1}
  \equiv
  \frac{\Phi\left(x_1+nh_1/2,x_2,x_3\right)
  +\Phi\left(x_1-nh_1/2,x_2,x_3\right)}
  {2}.
\end{equation}
}
In addition, the special interpolation operator of the product of two variables \tblue{$\Phi$ and $\Psi$} with stencil $n$ with respect to $x_1$ is defined as
\tblue{
\begin{align}
  \label{eq:op_interp_prod_n}
  \vwtilde{\Phi\Psi}{nx_1}
  \equiv&
  \frac{1}{2}
  \Phi\left(x_1+nh_1/2,x_2,x_3\right)
  \Psi\left(x_1-nh_1/2,x_2,x_3\right)
  \nonumber\\
  &+\frac{1}{2}
  \Phi\left(x_1-nh_1/2,x_2,x_3\right)
  \Psi\left(x_1+nh_1/2,x_2,x_3\right).
\end{align}
}
These operators are similarly defined in \torange{the} $x_2$- and $x_3$-directions.

The finite difference operator defined by Eq. (\ref{eq:op_diff_n}) is second-order accurate in space.
The higher-order finite difference operator can be represented by the weighted average of the $n$-stencil operator as
\tblue{
\begin{equation}
  \label{eq:op_diff}
  \frac{{\delta}{\Phi}}{{\delta}x_1}
  \equiv
  {\sum_{n=1}^{N_\mathrm{c}}}
  {c_n}\frac{{\delta_{2n}}{\Phi}}{{\delta_{2n}}x_1},
\end{equation}
}
where \tgreen{the coefficients $c_n$ satisfy} $\sum_nc_n=1$.
For example, the standard five-stencil \torange{fourth-order} accurate central difference operator \tblue{($N_c=2$, $c_1=4/3$, and $c_2=-1/3$)} can be expressed as
\begin{equation}
  \frac{{\delta}{\Phi}}{{\delta}x_1}
  =
  \frac{4}{3}
  \frac{{\delta_{2}}{\Phi}}{{\delta_{2}}x_1}
  -\frac{1}{3}
  \frac{{\delta_{4}}{\Phi}}{{\delta_{4}}x_1},
\end{equation}\tblue{
as shown in Morinishi et al. \cite{1998JCoPh.143...90M}.
}
These operators are similarly defined in other directions.
\tblue{
  In Sec. \ref{subsec:advection}, two specific implementation of the operator $\delta/\delta{x_j}$ are provided.
  We use the operator $\delta/\delta{x_j}$ as a building block of our formulation to simplify the notation.
}

The $n$-stencil finite difference operator satisfies the following three product rules:
\tblue{
\begin{equation}
  \label{eq:rule_prod_n}
  \Phi\frac{{\delta_{2n}}{\Psi}}{{\delta_{2n}}x_j}
  +\Psi\frac{{\delta_{2n}}{\Phi}}{{\delta_{2n}}x_j}
  =
  \frac{{\delta_{n}}\vwtilde{\Phi\Psi}{nx_j}}{{\delta_{n}}x_j},
\end{equation}
\begin{equation}
  \label{eq:rule_prod_na}
  \overline{\Phi}^{nx_j}\frac{{\delta_{n}}{\Psi}}{{\delta_{n}}x_j}
  +\overline{\Psi}^{nx_j}\frac{{\delta_{n}}{\Phi}}{{\delta_{n}}x_j}
  =
  \frac{{\delta_{n}}{\Phi\Psi}}{{\delta_{n}}x_j},
\end{equation}
}
and
\tblue{
\begin{equation}
  \label{eq:rule_prod_ns}
  \overline{
  \Psi
  \frac{{\delta_{n}}{\Phi}}{{\delta_{n}}x_j}
  }^{nx_j}
  +\Phi\frac{{\delta_{n}}{\Psi}}{{\delta_{n}}x_j}
  =
  \frac{{\delta_{n}}{\overline{\Phi}^{nx_j}}{\Psi}}{{\delta_{n}}x_j}.
\end{equation}
}
\torange{The right-hand sides} of these equations are conservative.
In this study, these product rules are frequently used to analyze the conservation properties of the finite difference scheme.
The conservation property of the higher-order operator also satisfies similar relations, such as
\tblue{
\begin{equation}
  \label{eq:rule_prod}
  \Phi\frac{\delta\Psi}{\delta{x_j}}
  +\Psi\frac{\delta\Phi}{\delta{x_j}}
  =\sum_{n=1}^{N_c}c_n
  \frac{{\delta_{n}}\vwtilde{\Phi\Psi}{nx_j}}{{\delta_{n}}x_j}.
\end{equation}
}
\torange{It is clear that the right-hand side} of the equation is conservative.

For future reference, we note the commutation rule of the finite difference operator expressed as
\tblue{
\begin{equation}
  \label{eq:rule_comm}
  \frac{{\delta_{n}}}{{\delta_{n}}x_i}
  \left(
  \frac{{\delta_{n}}{\Phi}}{{\delta_{n}}x_j}
  \right)
  =
  \frac{{\delta_{n}}}{{\delta_{n}}x_j}
  \left(
  \frac{{\delta_{n}}{\Phi}}{{\delta_{n}}x_i}
  \right).
\end{equation}
}
The higher-order operator also satisfies this rule.

\subsection{Energy-consistent formulation of \torange{the} compressible MHD equations}

We propose an energy-consistent formulation of the fully compressible magnetohydrodynamic equations expressed as:
\begin{equation}
  \label{eq:fdm_cont}
  \frac{\partial{\rho}}{\partial{t}}
  +\frac{\delta{\rho}{V_j}}{\delta{x_j}}
  =0,
\end{equation}
\begin{equation}
  \label{eq:fdm_mom}
  \frac{\partial{\rho}{V_i}}{\partial{t}}
  +\frac{1}{2}\frac{\delta{\rho}{V_j}{V_i}}{\delta{x_j}}
  +\frac{1}{2}{\rho}{V_j}\frac{\delta{V_i}}{\delta{x_j}}
  +\frac{1}{2}{V_i}\frac{\delta{\rho}{V_j}}{\delta{x_j}}
  =
  -\frac{\delta{P}}{\delta{x_i}}
  -B_j\left(
    \frac{\delta{B_j}}{\delta{x_i}}
    -\frac{\delta{B_i}}{\delta{x_j}}
  \right),
\end{equation}
\begin{equation}
  \label{eq:fdm_eint}
  \frac{\partial{e}}{\partial{t}}
  +\frac{\delta\left(e+P\right)V_j}{{\delta}{x_j}}
  =V_j\frac{\delta{P}}{\delta{x_j}},
\end{equation}
\begin{equation}
  \label{eq:fdm_induc}
  \frac{\partial{B_i}}{\partial{t}}
  +\frac{\delta\left({V_j}{B_i}-{V_i}{B_j}\right)}{\delta{x_j}}
  =0.
\end{equation}
\torange{Except the energy equation, the} hydrodynamic component of this formulation is similar to the second scheme proposed by Kok \cite{2009JCoPh.228.6811K}.
In Kok's scheme, the divergence of the enthalpy flux is expressed using the quadratic split operator.
We use the simple divergence form for the enthalpy flux in the internal energy equation for the consistency with the divergence of the mass flux in \tblue{Eq. (\ref{eq:fdm_cont})} and the \tgreen{curl} of the electric field in Eqs. (\ref{eq:fdm_induc}).
\torange{One advantage of our formulation is} that the pressure perturbation is not excited at the contact discontinuity where $P$, $e$, and $V_j$ are constant in space.
The energy conservation property is not affected by this change.
Note that the entropy conservation can also be achieved by changing the discretization of the equation of the continuity and the internal energy as shown by Kuya et al. \cite{2018JCoPh.375..823K}.
\torange{Given that} the purpose of this paper is to demonstrate the robustness of the energy-consistent scheme in stringent problems, the details of various formulations for the hydrodynamic equations \torange{are not discussed}.

\subsection{Momentum conservation in \torange{the} proposed formulation}
\label{subsec:form_mom}

The convective term in the equations of motion (Eqs. (\ref{eq:fdm_mom}))conserves the volume-averaged momentum as shown in the previous studies \cite{2009JCoPh.228.6811K,2010JCoPh.229..276M}.
Using the product rule (Eq. (\ref{eq:rule_prod})), the convective term can be rewritten as
\tblue{
\begin{equation}
  \frac{1}{2}\frac{\delta{\rho}{V_j}{V_i}}{\delta{x_j}}
  +\frac{1}{2}{\rho}{V_j}\frac{\delta{V_i}}{\delta{x_j}}
  +\frac{1}{2}{V_i}\frac{\delta{\rho}{V_j}}{\delta{x_j}}
  =\frac{1}{2}\frac{\delta{\rho}{V_j}{V_i}}{\delta{x_j}}
  +\sum_{n=1}^{N_c}\frac{c_n}{2}
  \frac{{\delta_{n}}\vwtilde{\left(\rho{V_j}\right)V_i}{nx_j}}
  {{\delta_{n}}x_j}.
\end{equation}
}
The \torange{right-hand side} is conservative, and the convective term conserves momentum.

The Lorentz force in the equations of motion can be also rewritten in the conservative form as
\tblue{
\begin{equation}
  \label{eq:fdm_fl}
  F^\mathrm{L}_i
  =
  -B_j\left(
    \frac{{\delta}{B_j}}{{\delta}{x_i}}
    -\frac{{\delta}{B_i}}{{\delta}{x_j}}
  \right)
  =
  -\sum_{n=1}^{N_c}c_n
  \left(
    \frac{1}{2}
    \frac{{\delta_{n}}{\vwtilde{B_jB_j}{nx_i}}}{{\delta_{n}}{x_i}}
    -\frac{{\delta_{n}}{\vwtilde{B_iB_j}{nx_j}}}{{\delta_{n}}{x_j}}
  \right)
  -B_i\frac{{\delta}{B_j}}{{\delta}{x_j}}.
\end{equation}
}
The last term \torange{disappears} when the numerical solenoidal condition
\begin{equation}
  \label{eq:fdm_divb}
  \frac{{\delta}{B_j}}{{\delta}{x_j}}
  =0
\end{equation}
is satisfied, i.e., the magnetic field divergence is zero.

\tgreen{By removing the $-B_i\divB$ term} in the \torange{right-hand side} of the momentum equations (see Eq. (\ref{eq:anl_fl_cnsv})), an energy-consistent formulation of the MHD equations \torange{can be constructed, which} conserves momentum even when the numerical solenoidal condition is violated.
However, this modification produces artificial forces parallel to the magnetic field lines \torange{as well as} an additional heating term $V_iB_i\divB$ in the internal energy equation to maintain the total energy conservation.
Therefore, this field-aligned artificial force may \torange{yield} an unphysical solution.
To avoid this problem, the above form of the Lorentz force is selected in this study.

\subsection{Total energy conservation in \torange{the} proposed formulation}

This subsection demonstrates the total energy conservation in the proposed formulation.
As discussed in previous studies, the skew-symmetric convective term conserves the kinetic energy with the appropriate discretization of the mass conservation law.
The convective term multiplied by $V_i$ can be written as
\tblue{
\begin{align}
  \label{eq:dfkin}
  V_i\left[
    \frac{1}{2}\frac{\delta{\rho}{V_j}{V_i}}{\delta{x_j}}
    +\frac{1}{2}{\rho}{V_j}\frac{\delta{V_i}}{\delta{x_j}}
    +\frac{1}{2}{V_i}\frac{\delta{\rho}{V_j}}{\delta{x_j}}
  \right]
  &=\frac{1}{2}V_i\frac{\delta{\rho}{V_j}{V_i}}{\delta{x_j}}
  +\frac{1}{2}{\rho}{V_j}{V_i}\frac{\delta{V_i}}{\delta{x_j}}
  +\frac{V_iV_i}{2}\frac{\delta{\rho}{V_j}}{\delta{x_j}}
  \nonumber\\
  &=
  \sum_{n=1}^{N_c}\frac{c_n}{2}
  \frac{{\delta_{n}}\vwtilde{\left(\rho{V_j}{V_i}\right)V_i}{nx_j}}
  {{\delta_{n}}x_j}
  +\frac{V_iV_i}{2}\frac{\delta{\rho}{V_j}}{\delta{x_j}}.
\end{align}
}
The kinetic energy equation is derived from the equations of motion Eqs. (\ref{eq:fdm_mom}) and the equation of continuity Eq. (\ref{eq:fdm_cont}) as
\tblue{
\begin{align}
  \label{eq:dtek}
  \frac{\partial{\rho{V_iV_i}/2}}{\partial{t}}
  &=V_i\frac{\partial{\rho{V_i}}}{\partial{t}}
  -\frac{V_iV_i}{2}\frac{\partial\rho}{\partial{t}}
  \nonumber\\
  &=
  -\sum_{n=1}^{N_c}\frac{c_n}{2}
  \frac{{\delta_{n}}\vwtilde{\left(\rho{V_j}{V_i}\right)V_i}{nx_j}}
  {{\delta_{n}}x_j}
  +W^\mathrm{P}+W^\mathrm{L}.
\end{align}
}
The first term of the last equation represents the divergence of the kinetic energy flux.
The second term of Eq. (\ref{eq:dfkin}) is canceled with the equation of continuity.

The work done by the pressure gradient force
\begin{equation}
  W^\mathrm{P}
  =-V_j\frac{{\delta}{P}}{{\delta}{x_j}}
\end{equation}
represents the interaction between the internal and kinetic energy equations.
Apparently, this is exactly equivalent to the \torange{right-hand side} of the internal energy equation Eq. (\ref{eq:fdm_eint}), apart from the difference in sign.

The magnetic energy equation is derived by multiplying the induction equation Eq. (\ref{eq:fdm_induc}) by $B_i$, such that
\tblue{
\begin{align}
  \frac{{\partial}{B_iB_i/2}}{{\partial}{t}}
  &=
  B_i\frac{{\partial}{B_i}}{{\partial}{t}}
  =
  -B_i\frac{\delta\left({V_j}{B_i}-{V_i}{B_j}\right)}{\delta{x_j}}
  \nonumber\\
  &=
  -\sum_{n=1}^{N_c}c_n
  \frac{{\delta_{n}}{
    \vwtilde{B_i\left({V_j}{B_i}-{V_i}{B_j}\right)}{nx_i}}
  }{{\delta_{n}}{x_i}}
  -W^\mathrm{L},
\end{align}
}
where the first term in the last equation is (minus of) the divergence of the Poynting flux and the second term represents the work done by the Lorentz force
\begin{equation}
  W^\mathrm{L}
  =-V_iB_j\left(
    \frac{{\delta}{B_j}}{{\delta}{x_i}}
    -\frac{{\delta}{B_i}}{{\delta}{x_j}}
  \right).
\end{equation}
Comparison with the discretized Lorentz force of Eq. (\ref{eq:fdm_fl}) reveals that the interaction between the kinetic and magnetic energy equations is consistent in the discretization level.
Note that we did not use the solenoidal condition of Eq. (\ref{eq:fdm_divb}) in the derivation of the magnetic energy equation.
Thus, the proposed formulation conserves the total energy even when the $\divB=0$ condition is violated.

\subsection{Energy-consistent formulation of \torange{the} nonlinear filtering flux}
\label{subsec:ec_filtering}

\torange{Shock waves and discontinuities are often involved in the} solutions of compressible hydrodynamics and magnetohydrodynamics.
In the approach adopted in this work, the numerical simulation is stabilized by diffusing the sharp gradient through \torange{introducing} of the nonlinear filtering flux in the semi-discrete formulation of \tgreen{Eqs. (\ref{eq:fdm_cont})--(\ref{eq:fdm_induc})}.
An energy-consistent formulation is obtained even when this nonlinear filtering flux is introduced.

We design an energy-consistent formulation for the filtering flux defined at the center of the cell face.
Note that many linear and nonlinear filtering schemes can be reduced to this formulation.
The filtering process of the variable $U$ can be expressed as
\begin{equation}
  \frac{{\partial}{U}}{{\partial}{t}}
  =
  \frac{{\delta_1}{D_j(U)}}{{\delta_1}{x_j}},
\end{equation}
where $D_j(U)$ is the filtering flux in the $j$-th direction defined at the cell edge.
The typical first-order filtering flux can be expressed as
\begin{equation}
  D_j(U)={\eta(U)}\frac{{\delta_1}{U}}{{\delta_1}{x_j}},
\end{equation}
where $\eta(U)$ is the diffusion coefficient of $U$.
The construction of the higher-order filtering flux is described in Section \ref{sec:implement}.
\tblue{
The energy-consistent property is not affected by the specific implementation of the filtering flux $D_j(U)$ as long as the filtering flux is defined at the cell face.
}
In this study, we apply this filtering scheme on the mass density $\rho$, the momentum density $\rho{V_i}$, the magnetic flux density $B_i$, and the internal energy density $e$.

When the filtering scheme is applied on the equations of motion or the induction equations, the viscous heating $Q_\mathrm{vis}$ or the resistive heating $Q_\mathrm{res}$ should be included as heating terms in the internal energy equation to maintain the total energy conservation.
We propose the energy-consistent discretization of the viscous and resistive heating rates corresponding to the filtering flux as
\begin{equation}
  \label{eq:def_qvis}
  Q_\mathrm{vis}
  =
  \overline{D_j({\rho}V_i)
  \frac{{\delta_1}{V_i}}{{\delta_1}{x_j}}
  }^{1x_j}
  -\overline{
  \frac{D_j(\rho)}{2}
  \frac{{\delta_1}{V_iV_i}}{{\delta_1}x_j}
  }^{1x_j}
\end{equation}
and
\begin{equation}
  \label{eq:def_qres}
  Q_\mathrm{res}
  =
  \overline{D_j(B_i)
  \frac{{\delta_1}{B_i}}{{\delta_1}{x_j}}}^{1x_j},
\end{equation}
respectively.
These expressions can be rewritten as
\begin{equation}
  Q_\mathrm{vis}
  =
  \frac{{\delta_1}}{{\delta_1}{x_j}}
  \left[
  \overline{V_i}^{1x_j}{D_j({\rho}V_i)}
  -\frac{1}{2}\overline{V_iV_i}^{1x_j}{D_j(\rho)}
  \right]
  -V_i\frac{{\delta_1}{D_j({\rho}V_i)}}{{\delta_1}{x_j}}
  +\frac{V_iV_i}{2}\frac{{\delta_1}{D_j(\rho)}}{{\delta_1}{x_j}}
\end{equation}
and
\begin{equation}
  Q_\mathrm{res}
  =
  \frac{{\delta_1}\overline{B_i}^{1x_j}{D_j(B_i)}}{{\delta_1}{x_j}}
  -B_i\frac{{\delta_1}{D_j(B_i)}}{{\delta_1}{x_j}}.
\end{equation}
Here, the first terms in these two equations are the kinetic and magnetic energy transport by the filtering fluxes, respectively.
The other terms correspond to the time variation of the kinetic and magnetic energies produced by the filtering flux.
\tgreen{Clearly}, these heating rates are energy-consistent in terms of the exchange among the internal, kinetic, and magnetic energies.

The viscous and resistive heating rates defined in Eqs. (\ref{eq:def_qvis}) and (\ref{eq:def_qres}) become positive under the specific condition.
We assume the filtering flux of the momentum density ${\rho}V_i$ evaluated from those of the mass density and velocity field as
\begin{equation}
  \label{eq:def_dmom}
  D_j({\rho}V_i)=
  \overline{\rho}^{1x_j}D_j(V_i)
  +\overline{V_i}^{1x_j}D_j(\rho).
\end{equation}
Then, the viscous heating rate Eq. (\ref{eq:def_qvis}) can be rewritten as
\begin{equation}
  Q_\mathrm{vis}
  =
  \overline{
    \left[
      \overline{\rho}^{1x_j}D_j(V_i)
      +\overline{V_i}^{1x_j}D_j(\rho)
    \right]
    \frac{{\delta_1}{V_i}}{{\delta_1}{x_j}}
  }^{1x_j}
  -\overline{
  \frac{D_j(\rho)}{2}
  \frac{{\delta_1}{V_iV_i}}{{\delta_1}x_j}
  }^{1x_j}
  =\overline{
    \overline{\rho}^{1x_j}D_j(V_i)
    \frac{{\delta_1}{V_i}}{{\delta_1}{x_j}}
  }^{1x_j},
\end{equation}
where the product rule \torange{of} Eq. (\ref{eq:rule_prod_na}) for \tblue{$\Phi=\Psi=V_i$} \torange{is used}.
The above viscous heating rate becomes positive if the sign of $D_j(V_i)$ is \torange{the} same as that of $\delta_1V_i/\delta_1x_j$.
This condition is often satisfied considering that the $D_j(V_i)$ is the filtering flux acting to diffuse $V_i$ in \torange{the} $x_j$-direction.
Similarly, the resistive heating rate Eq. (\ref{eq:def_qres}) becomes positive when $D_j(B_i)$ and $\delta_1B_i/\delta_1x_j$ have the same sign.

\subsection{Treatment of magnetic monopole}
\label{subsec:divb}

In the multi-dimensional simulations of the magnetohydrodynamics, the solenoidal condition $\divB=0$ should be preserved throughout the time integration.
\torange{Violating} the solenoidal rule may produce artificial inconsistencies in the numerical solution.

\torange{The} central difference scheme \torange{is known to} produce no numerical magnetic monopole \cite{2000JCoPh.161..605T} as
\begin{align}
  \frac{{\delta}}{{\delta}{x_i}}
  \left(\frac{{\partial{B_i}}}{{\partial}{t}}\right)
  &=
  -\frac{\delta}{\delta{x_i}}
  \left[
  \frac{\delta\left({V_j}{B_i}
  -{V_i}{B_j}\right)}{\delta{x_j}}
  \right]
  \nonumber\\
  &=
  -\frac{{\delta}}{{\delta}{x_i}}
  \left(
  \frac{{\delta}{V_j}{B_i}}{{\delta}{x_j}}
  \right)
  +\frac{{\delta}}{{\delta}{x_i}}
  \left(
  \frac{{\delta}{V_i}{B_j}}{{\delta}{x_j}}
  \right)
  \nonumber\\
  &=0.
\end{align}
Here, we use the commutation rule of the central difference operators (Eq. (\ref{eq:rule_comm})).
On the other hand, the filtering flux violates the solenoidal condition $\divB=0$ as
\begin{equation}
  \frac{{\delta}}{{\delta}{x_k}}
  \left[
  \frac{{\delta_{1}}D_j(B_i)}{{\delta_{1}}{x_j}}
  \right]
  \neq0.
\end{equation}
We need some treatment to control the magnetic monopole on the numerical solution.

In this study, the \tgreen{hyperbolic/parabolic} divergence cleaning method \cite{2002JCoPh.175..645D} \torange{is used} to control the divergence of the magnetic field.
The numerical magnetic monopole \torange{is} transported and diffused by introducing an additional equation of the \tgreen{generalized Lagrange multiplier (GLM)} $\psi$ as
\begin{equation}
  \frac{{\partial}\psi}{{\partial}{t}}
  =
  -c_{\psi}^2\frac{{\delta}B_i}{{\delta}{x_i}}
  -\frac{\psi}{\tau_{\psi}},
\end{equation}
where the propagation speed $c_\psi$ and the damping rate $\tau_\psi$ are the tunable parameters that control the efficiency of the divergence cleaning.
Correspondingly, the induction equations are corrected as
\begin{equation}
  \frac{{\partial}B_i}{{\partial}{t}}
  =
  -\frac{{\delta}\psi}{{\delta}{x_i}},
\end{equation}
\tgreen{where, for the sake of clarity, we dropped the terms containing the curl} of the electric field and the filtering flux on the magnetic flux density.
\torange{It is clear that the} divergence cleaning method conserves the magnetic flux in a volume.

Although the \tgreen{hyperbolic/parabolic} divergence cleaning method can efficiently reduce the divergence of the magnetic field, $\divB$ still remains in the numerical solution.
The numerical discretization should be energy-consistent even when the small numerical error of $\divB$ exists.
This method produces the additional term in the magnetic energy equation as
\tblue{
\begin{equation}
  \frac{{\partial}B_i^2/2}{{\partial}{t}}
  =
  -B_i\frac{{\delta}\psi}{{\delta}{x_i}}
  =
  -\sum_{n=1}^{N_c}c_n
  \frac{{\delta_{n}}\vwtilde{{B_i}{\psi}}{nx_i}}{{\delta_{n}}{x_i}}
  -Q_\psi,
\end{equation}
}
where we define the heating term due to the divergence cleaning method
\begin{equation}
  \label{eq:def_qpsi}
  Q_\psi=-\psi\frac{{\delta}B_i}{{\delta}{x_i}}.
\end{equation}
In the steady state, this term becomes positive, as
\begin{equation}
  Q_\psi
  \rightarrow
  \tau_{\psi}c_{\psi}^2
  \left(\frac{{\delta}B_i}{{\delta}{x_i}}\right)^2
  \geq0.
\end{equation}
Introduction of $Q_\psi$ in the internal energy equation provides the energy-consistent formulation with the \tgreen{hyperbolic/parabolic} divergence cleaning.

\subsection{Summary of \torange{the} proposed scheme}
\label{subsec:fdm_final}

The energy-consistent formulation of the compressible MHD equations is summarized as follows:
\begin{equation}
  \label{eq:final_cont}
  \frac{\partial{\rho}}{\partial{t}}
  +\frac{\delta{\rho}{V_j}}{\delta{x_j}}
  =\frac{{\delta_{1}}{D_j(\rho)}}{{\delta_{1}}{x_j}},
\end{equation}
\begin{equation}
  \label{eq:final_mom}
  \frac{\partial{\rho}{V_i}}{\partial{t}}
  +\frac{1}{2}\frac{\delta{\rho}{V_j}{V_i}}{\delta{x_j}}
  +\frac{1}{2}{\rho}{V_j}\frac{\delta{V_i}}{\delta{x_j}}
  +\frac{1}{2}{V_i}\frac{\delta{\rho}{V_j}}{\delta{x_j}}
  +\frac{\delta{P}}{\delta{x_i}}
  +B_j\frac{\delta{B_j}}{\delta{x_i}}
  -B_j\frac{\delta{B_i}}{\delta{x_j}}
  =\frac{{\delta_{1}}{D_j(\rho{V_i})}}{{\delta_{1}}{x_j}},
\end{equation}
\begin{equation}
  \label{eq:final_eint}
  \frac{\partial{e}}{\partial{t}}
  +\frac{\delta\left(e+P\right)V_j}{{\delta}{x_j}}
  =-V_j\frac{\delta{P}}{\delta{x_j}}
  +\frac{{\delta_{1}}{D_j(e)}}{{\delta_{1}}{x_j}}
  +Q_\mathrm{vis}+Q_\mathrm{res}+Q_\psi,
\end{equation}
\begin{equation}
  \label{eq:final_induc}
  \frac{\partial{B_i}}{\partial{t}}
  +\frac{\delta\left({V_j}{B_i}-{V_i}{B_j}\right)}{\delta{x_j}}
  +\frac{\delta\psi}{\delta{x_i}}
  =\frac{{\delta_{1}}{D_j(B_i)}}{{\delta_{1}}{x_j}},
\end{equation}
\begin{equation}
  \label{eq:final_psi}
  \frac{{\partial}\psi}{{\partial}{t}}
  +c_{\psi}^2\frac{{\delta}B_j}{{\delta}{x_j}}
  =
  -\frac{\psi}{\tau_{\psi}}
  +\frac{{\delta_{1}}{D_j(\psi)}}{{\delta_{1}}{x_j}}.
\end{equation}
The viscous and resistive heating rates \torange{($Q_\mathrm{vis}$ and $Q_\mathrm{res}$)} are defined in Eqs. (\ref{eq:def_qvis}) and \tblue{(\ref{eq:def_qres}), respectively}.
The heating caused by the numerical magnetic monopole \torange{$Q_\psi$} is defined in Eq. (\ref{eq:def_qpsi}).
In this formulation, the internal, kinetic, and magnetic energy equations are consistent even if the nonlinear filtering flux and the divergence cleaning \torange{are} introduced.
The energy-consistent formulation of the compressible hydrodynamic equations can be derived by setting $B_i$, $\psi$, $Q_\mathrm{res}$, and $Q_\psi$ to be zero in Eqs. (\ref{eq:final_cont})--(\ref{eq:final_eint}).
The proposed formulation is independent of the specific implementations of the temporal discretization, the central difference operator $\delta/\delta{x_j}$, and the nonlinear filtering flux $D_j(U)$ for each variable $U$.
The details of our implementation are described in Section \ref{sec:implement}.

\section{Numerical implementation}
\label{sec:implement}

This section describes the details of our implementation of the proposed formulation.
In the proposed formulation Eqs. (\ref{eq:final_cont})-(\ref{eq:final_psi}), the time derivative is not discretized.

\subsection{Base schemes for scalar advection equation}
\label{subsec:advection}

In this subsection, we describe our implementation and performance of the base schemes using the one-dimensional scalar advection equation:
\begin{equation}
  \frac{{\partial}{U}}{{\partial}{t}}
  +a\frac{{\partial}{U}}{{\partial}{x_1}}=0,
  \label{eq:anl_1d_adv}
\end{equation}
where $U$ is the scalar variable and $a$ is the advection speed.
The semi-discrete form of the above equation can be written as
\begin{equation}
  \frac{{\partial}{U}}{{\partial}{t}}
  +a\frac{{\delta}{U}}{{\delta}{x_1}}
  =\frac{{\delta_1}D_1(U)}{{\delta_1}{x_1}}.
  \label{eq:semi_1d_adv}
\end{equation}
We implemented \torange{the} second- and fifth-order schemes to show the general performance of the proposed formulation.
\tblue{
  The important feature of these schemes is the small numerical overshoot, which seems to contribute the high robustness of our formulation shown in Sec. \ref{sec:test}.
  We speculate that the similar high robustness can be achieved with other combination of the central difference operator and filtering flux, as long as the produced numerical overshoot is sufficiently small.
}

In the second-order scheme, we used the second-order central difference operator $\delta/\delta{x_j}$ defined in Eq. (\ref{eq:op_diff}) by setting \tblue{$N_\mathrm{c}=1$ and $c_1=1$}.
The second-order filtering flux was taken from the Jameson-Schmidt-Turkel (JST) scheme \cite{1981fpd..conf.....J,jameson2017origins}.
The JST scheme is total variation diminishing if the limiter function is chosen appropriately \cite{swanson1992central}.
Among the variant of the JST scheme, we used the symmetric limited positive (SLIP) scheme \cite{1995IJCFD...4..171J}, which is proven to be positivity preserving for the scalar conservation law.
The filtering flux of the SLIP scheme for the scalar advection equation can be written as
\begin{equation}
  D_1^{\rm JST}(U;x_1+h_1/2)
  =\frac{\left|a\right|}{2}
  \left[
    \Delta{U}_{x_1+h_1/2}
    -\mathcal{L}\left(
      \Delta{U}_{x_1+3h_1/2},
      \Delta{U}_{x_1-h_1/2}
    \right)
  \right],
\end{equation}
where $\Delta{U}_{x}=U(x+h_1/2)-U(x-h_1/2)$ is the difference of $U$ between two adjacent cells and $\mathcal{L}(a,b)$ is the limiter function.
\tblue{
  We used van Leer's limiter (originaly introduced in \cite{1974JCoPh..14..361V}, implemented as described by \cite{1995IJCFD...4..171J}) for the limiter function $\mathcal{L}$.
}


To demonstrate the applicability of the proposed scheme in the \torange{higher order}, we also implemented the fifth-order scheme using the sixth-order central difference operator $\delta/\delta{x_j}$ for \tblue{$N_\mathrm{c}=3$, $c_1=3/2$, $c_2=-3/5$, and $c_3=1/10$}.
The filtering flux is based on the fifth-order weighted essentially non-oscillatory (WENO) reconstruction \cite{1996JCoPh.126..202J}.
The WENO scheme was originally developed in the context of the upwind scheme and is not intended to be used as the filtering flux for the central difference method.
The filtering flux \torange{in the $x_1$-direction} based on the WENO reconstruction can be written as
\begin{equation}
  D_1^{\rm WENO}(U;x_1+h_1/2)
  =\frac{c_{\rm diff}\left|a\right|}{2}
  \left[
    U^\mathrm{R}(x_1+h_1/2)-U^\mathrm{L}(x_1+h_1/2)
  \right],
\end{equation}
where $U^\mathrm{R}$ and $U^\mathrm{L}$ are the upwind value of $U$ reconstructed from the positive and negative $x_1$-direction, respectively.
A positive diffusion parameter $c_{\rm diff}$ \torange{of order unity} is introduced to control the numerical diffusion by the filtering flux.

\tblue{
The one-dimensional scalar advection equation Eq. (\ref{eq:anl_1d_adv}) conserves its energy $U^2/2$ as
\begin{equation}
  \frac{{\partial}{U^2/2}}{{\partial}{t}}
  +a\frac{{\partial}{U^2/2}}{{\partial}{x_1}}=0.
\end{equation}
The corresponding semi-discrete equation can be obtained by multiplying Eq. (\ref{eq:semi_1d_adv}) by $U$ as
\begin{equation}
  U\frac{{\partial}{U}}{{\partial}{t}}
  +a\left(
    \sum_{n=1}^{N_c}c_n
    \frac{{\delta_{n}}\vwtilde{U^2/2}{nx_1}}{{\delta_{n}}x_1}
  \right)
  -\frac{{\delta_1}\overline{U}^{1x_1}D_1(U)}{{\delta_1}{x_1}}
  =
  -\overline{D_1(U)\frac{{\delta_1}U}{{\delta_1}{x_1}}}^{1x_1},
  \label{eq:semi_1d_adv_ekin}
\end{equation}
where the right-hand side represents the energy loss by the filtering flux $D_1(U)$.
Note that the similar terms of the energy loss in the kinetic and magnetic energy equations in our MHD formulation are treated as the viscous and resistive heating rates in the internal energy equation to achieve the total energy conservation even if the filtering flux is introduced (see Sec. \ref{subsec:ec_filtering}).
The semi-discrete scheme Eq. (\ref{eq:semi_1d_adv}) conserves the energy $U^2/2$ if the energy loss by the filtering flux is negligible.
However, the temporal discretization can violate the energy conservation because the analytical identity $U\partial{U}/\partial{t}=\partial{(U^2/2)}\partial{t}$ cannot be satisfied by a time integration scheme in general, like most of the Runge-Kutta methods.
One simple exception is the implicit midpoint method, which is symmetric in temporal direction.
Similar to the advection equation, the total energy conservation in the semi-discrete formulation of the MHD equations in Eqs. (\ref{eq:final_cont})--(\ref{eq:final_psi}) produces the conservation error of the total energy by the time integration scheme.
More detailed descriptions on the relation between the energy conservation and the temporal integration method can be found in \cite{2009JCoPh.228.6811K,2010JCoPh.229..276M}.
}

\tblue{
In all test problems described in this paper, we used the third-order optimal strongly stability preserving (SSP) Runge-Kutta method \cite{1988JCoPh..77..439S} for the SSP property in the nonlinear equations \cite{gottlieb2009high} and its frequent use with the fifth-order WENO scheme \cite{1996JCoPh.126..202J}.
We also found that the four-step second-order Runge-Kutta method by Jameson \& Baker \cite{jameson1983solution} also performs well even in the stringent problems in Secs. \ref{subsec:blast} and \ref{subsec:rotor}.
Therefore, we speculate that the robustness of the proposed formulation may be weakly affected by the choice of the time integration scheme.
}

\begin{figure}[!ht]
  \centering
  \plotone{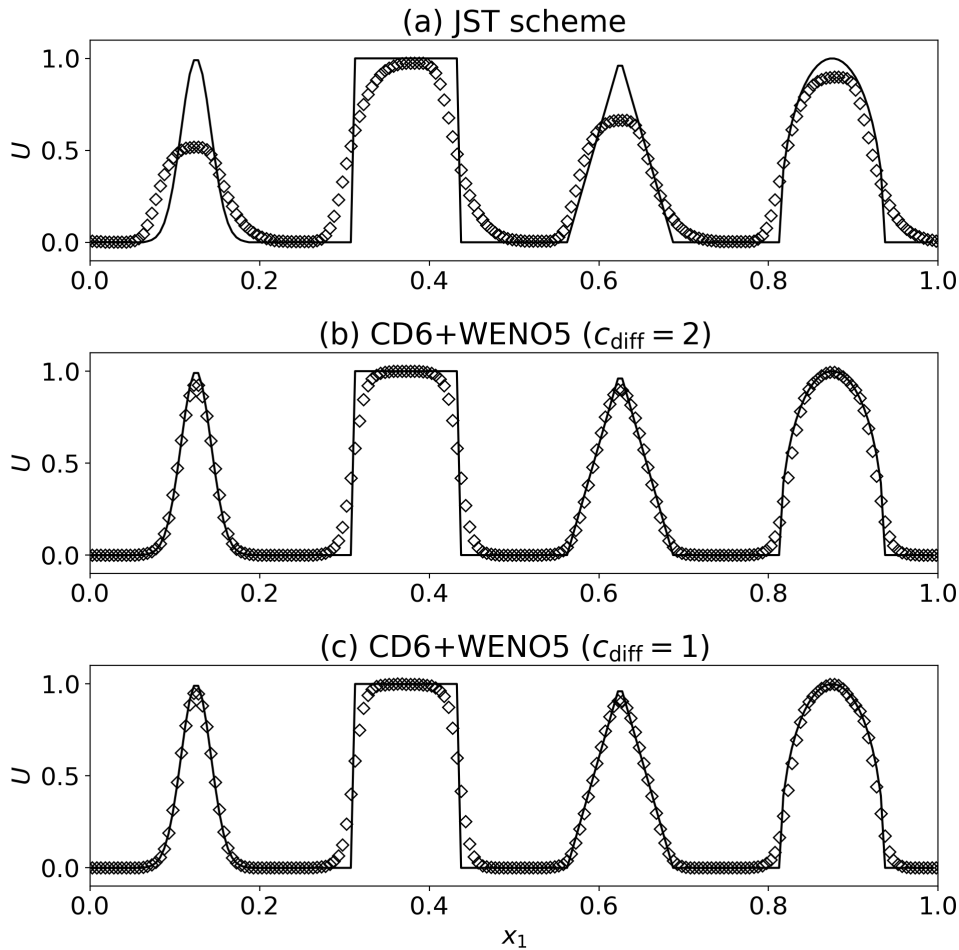}
  \caption{
  Results of one-dimensional scalar advection problem using 200 \tblue{grid points}.
  Shown are the results at $t=1$ for
  (a) JST scheme,
  (b) sixth-order central difference method
  with the filtering flux by \torange{the} WENO5 scheme with $c_{\rm diff}=2$, and
  (c) sixth-order central difference method
  with the filtering flux by \torange{the} WENO5 scheme with $c_{\rm diff}=1$.
  The solid lines indicate the analytical solution.
  The diamond symbols indicate the numerical results.
  The Courant number of \tred{0.3} is used.
  \label{fig:adv1d_full}
  }
\end{figure}
Figure \ref{fig:adv1d_full} shows the solution of the one-dimensional scalar advection problem after one period.
The initial profile is identical to that of the test problem in Suresh \& Huynh \cite{1997JCoPh.136...83S}.
The results show that both the second-order scheme with the JST filtering flux and the sixth-order scheme with the fifth-order WENO filtering flux efficiently \torange{preserve} the monotonicity of the solution.

\begin{figure}[!ht]
  \centering
  \plotone{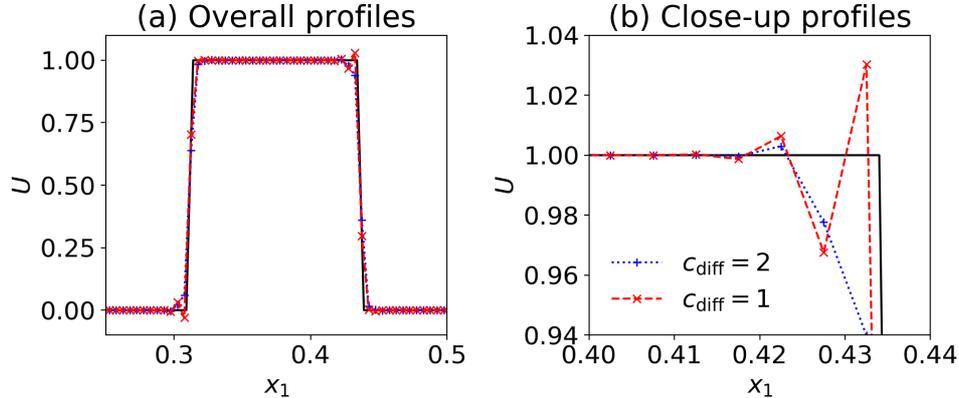}
  \caption{
  One-dimensional scalar advection problem using 200 \tblue{grid points}
  with \torange{sixth-order} central difference method
  with the filtering flux by \torange{the} WENO5 scheme.
  Shown are the results one step after the integration (at $t=0.0015$)
  for $c_{\rm diff}=2$ (blue dotted lines with plus)
  and $c_{\rm diff}=1$ (red dashed lines with cross).
  The solid lines indicate the analytical solution.
  The Courant number of \tred{0.3} is used.
  \label{fig:adv1d_step}
  }
\end{figure}
The dependence on the diffusion parameter $c_{\rm diff}$ of the WENO filter becomes clear near the discontinuity.
Figure \ref{fig:adv1d_step} shows solutions for the same problem as Fig. \ref{fig:adv1d_full}, but showing one step after the time integration (at $t=0.0015$).
Small numerical overshoots are found near the discontinuity if $c_{\rm diff}=1$.
We found that the overshoot can be efficiently damped by doubling the numerical diffusion in the WENO filter ($c_{\rm diff}=2$).
However, small overshoot of about $10^{-5}$ still remains even if $c_{\rm diff}=2$.
\tred{
  We note that this value of $c_{\rm diff}=2$ is not guaranteed theoretically nor optimized for general problems.
  However, this choice was found to show high numerical stability in all test problems presented in Sec. \ref{sec:test}.
}
Because the formulation proposed in this study is independent of the specific implementation of the filtering flux, better methods of evaluating the filtering flux, \tred{especially with less numerical overshoot and higher-order spatial convergence, should be investigated} in \torange{future studies}.

\subsection{Filtering flux for system equations}
\label{subsec:filter}

For the primitive variables in the MHD equations with the \tgreen{hyperbolic/parabolic} divergence cleaning method, \tblue{$W_i=(\rho,e,V_1,V_2,V_3,B_1,B_2,B_3,\psi)$}, the one-dimensional form of the evolution equations can be written as
\begin{equation}
  \frac{\partial{W_l}}{\partial{t}}
  +\sum_{n}A_{ln}\frac{\partial{W_n}}{\partial{x_1}}
  =0,
\end{equation}
where $A$ is the coefficient matrix for the primitive variables.
The matrix $A$ can be diagonalized as
\begin{equation}
  A_{ln}=\sum_{m}R_{lm}\lambda_{m}L_{mn},
\end{equation}
where $L$ and $R$ are the left and right eigenmatrices \cite{1998ApJS..116..119B,2008ApJS..178..137S}, respectively, and $\lambda_m$ \torange{indicates} the eigenvalues of the matrix $A$.

There are several ways \torange{to extend} the filtering flux for the scalar advection equation to the system equations.
\torange{
  Here, we implemented two methods, namely, the LLF-type and the Roe-type filtering schemes.
}
\torange{The first} approach to evaluate the filtering flux is \torange{to diffuse} all variables with the same characteristic speed, as in the local Lax-Friedrichs (LLF) scheme \cite{lax1954weak,2000JCoPh.160..241K}.
The filtering flux of the primitive variables in the $x_1$-direction at the location $x_1+h_1/2$ is computed as
\begin{equation}
  D(W_i;x_1+h_1/2)=\frac{a_\mathrm{max}}{2}
  \mathcal{D}\left[
    {W}({x_1+h_1/2{\pm}h_1/2}),
    {W}({x_1+h_1/2{\pm}3h_1/2}),
    \cdots
  \right],
\end{equation}
where $a_\mathrm{max}=\max\left(\left|\lambda_1\right|,...,\left|\lambda_9\right|\right)$ is the maximum phase speed of the waves in the system and the discrete operator $\mathcal{D}$ represents the limiting process in the JST and WENO5 filtering \torange{fluxes}.
The filtering flux for the momentum density was computed from that of the mass density and the velocity field using Eq. (\ref{eq:def_dmom}).

\torange{The second} filtering flux \torange{implemented in this study} is based on \torange{Roe's approximate Riemann solver} \cite{roe1981approximate,swanson1992central,1999JCoPh.150..199Y}, \torange{which is} given by
\begin{equation}
  D(W_i;x_1+h_1/2)=
  \frac{1}{2}\sum_mR_{lm}\left|\lambda_{m}\right|
  \mathcal{D}\left[
    Q_m(x_1+h_1/2{\pm}h_1/2),
    Q_m(x_1+h_1/2{\pm}3h_1/2),
    \cdots
  \right],
\end{equation}
where
\begin{equation}
  Q_m(x_1)=\sum_nL_{mn}{W_n(x_1)}
\end{equation}
is the characteristic variables with the phase speed of $\lambda_m$.
The left and right eigenmatrices $R_{ij}$ \torange{and} $L_{jk}$ and the eigenvalues $\lambda_j$ \torange{are} evaluated at the fixed location $x_1+h_1/2$.
\tgreen{
In our implementation, these eigenmatrices and eigenvalues are derived from the arithmetic average of the primitive variables at the adjacent grids, i.e., $\overline{W_i}^{1x_1}(x_1+h_1/2)=(W_i(x_1)+W_i(x_1+h_1))/2$.
Although the original Roe's scheme \cite{roe1981approximate} employs the Roe-average, the arithmetic average has often been utilized in practical problems, and it seems to work well \cite{1995ApJ...442..228R}.
}
We also used the first entropy fix suggested by Harten \& Hyman \cite{1983JCoPh..50..235H} to evaluate the eigenvalues.
The filtering flux of the momentum density was derived using Eq. (\ref{eq:def_dmom}).

In the Roe-type method, the characteristic variables are used \torange{for the filtered variables} instead of the primitive variables to improve the monotonicity of the numerical solution \cite{1997JCoPh.131....3H}.
The positivity of the viscous and resistive heating rates discussed in \torange{Sec.} \ref{subsec:ec_filtering} may be violated when the Roe-type method is applied.
\tgreen{
We actually found that the Roe-type method provides less robustness than the LLF-type method when the internal energy is much smaller than the kinetic or magnetic energy.
}

\tred{
\subsection{Time step size and parameters of hyperbolic/parabolic divergence cleaning method}
}
\label{subsec:cfl}

\tred{
The time step size $\varDelta{t}$ is determined from the Courant-Friedrichs-Lewy (CFL) condition
\begin{equation}
  \varDelta{t}=
  \min_{x_1,x_2,x_3}
  \left\{
    \sigma_\mathrm{CFL}
    \left[
      \sum_{i=1}^{N_\mathrm{D}}
      \frac{\max(\left|\lambda_{i,2}(x_1,x_2,x_3)\right|,
      \left|\lambda_{i,8}(x_1,x_2,x_3)\right|)}{h_i}
    \right]^{-1}
  \right\},
\end{equation}
where $\lambda_{i,2}$ and $\lambda_{i,8}$ are the eigenvalues for the fast magneto-acoustic waves calculated for the $x_i$-direction, $\sigma_\mathrm{CFL}$ is the Courant number of order unity, and $N_\mathrm{D}=1,2,3$ is the number of dimensions.
In the most test problems shown in Sec. 4, we set the default Courant number $\sigma_\mathrm{CFL}=0.3$. Although most problems can be solved with larger $\sigma_\mathrm{CFL}$ like $0.4$ or $0.5$, the value of $0.3$ is required for stability in the stringent problems described in Secs. \ref{subsec:blast} and \ref{subsec:rotor}.
}

\tred{
In the multi-dimensional problems, we used the hyperbolic/parabolic divergence cleaning method.
The propagation speed of the magnetic field divergence $c_\psi$ is computed from the CFL condition at each time step as
\begin{equation}
  c_\psi=\sigma_\mathrm{CFL}
  \left[
    \varDelta{t}\sum_{i=1}^{N_\mathrm{D}}\frac{1}{h_i}
  \right]^{-1}.
\end{equation}
The damping time $\tau_\psi$ was set to $10\varDelta{t}$.
More detailed discussion on the choice of the damping time $\tau_\psi$ can be found in \cite{2002JCoPh.175..645D,2010JCoPh.229.5896M}.
}

\section{Test problems}
\label{sec:test}


Based on the newly proposed formulation described in \torange{Sec.} \ref{subsec:fdm_final}, four schemes were implemented, \torange{namely,} JST-LLF, JST-Roe, WENO5-LLF, and WENO5-Roe schemes.
In the JST-LLF and JST-Roe schemes, the second-order central difference with the filtering flux based on the SLIP version of the Jameson-Schmidt-Turkel scheme \torange{was} used.
In the two WENO5 schemes, the sixth-order central difference with the filtering flux based on the fifth-order WENO reconstruction is used.
The JST-LLF and WENO5-LLF schemes are \torange{based on the LLF-type filtering scheme,} and the JST-Roe and WENO5-Roe schemes are implemented with {the Roe-type filtering scheme}.
See Section \ref{sec:implement} for the details of our implementation.
We found that the LLF-type \torange{filtering scheme} becomes more numerically robust than the Roe-type \torange{scheme}.
Most of the test problems presented in this study can be solved with both the LLF- and Roe-type \torange{filtering schemes}, unless otherwise noted.
The exceptions are the most stringent, non-standard test problems described in \torange{Secs.} \ref{subsec:blast} and \ref{subsec:rotor}.

All four schemes were integrated using the three-step third-order optimal SSP Runge--Kutta method \cite{1989JCoPh..83...32S}.
The Courant number \torange{$\sigma_\mathrm{CFL}$} of \tred{$0.3$} was used \torange{in all test problems,} unless otherwise noted.
The equation of states of ideal gas was used in all problems.

\tgreen{
\subsection{Two-dimensional magnetized iso-density vortex problem}
}

\begin{table}[!ht]
  \caption{
    \tgreen{
  $L_1$ norm errors and corresponding convergence rates for the two-dimensional magnetized iso-density vortex problem.
  The errors are measured in the $x_1$-component of the magnetic field $B_1$ and the divergence of the magnetic field $\divB$.
  Note that we set $q=0.5$ for the second-order schemes (JST-Roe, JST-LLF, and CD2) and $q=1.0$ for the higher-order schemes (WENO5-Roe, WENO5-LLF, and CD6).
    }
  }
  \label{table:vortex}
  \center
  \fontsize{9pt}{9pt}\selectfont
  \begin{tabular}[tb]{cccccc}
    \hline\rule{0pt}{2.6ex}
    Method& Number of mesh&
    $L_1$ error of $B_1$& $L_1$ order of $B_1$&
    $L_1$ error of $\divB$& $L_1$ order of $\divB$\\
    \hline\rule{0pt}{2.6ex}
    JST-LLF&${32}\times{32}$&$\num{1.0526E-02}$&&$\num{4.0819E-04}$&\\
    &${64}\times{64}$&$\num{5.1003E-03}$&$\num{1.05}$&$\num{4.1512E-04}$&$\num{-0.02}$\\
    &${128}\times{128}$&$\num{1.5464E-03}$&$\num{1.72}$&$\num{1.8711E-04}$&$\num{1.15}$\\
    &${256}\times{256}$&$\num{4.4229E-04}$&$\num{1.81}$&$\num{4.3809E-05}$&$\num{2.09}$\\
    &&&&&\\
    JST-Roe&${32}\times{32}$&$\num{7.7712E-03}$&&$\num{5.2883E-04}$&\\
    &${64}\times{64}$&$\num{2.9609E-03}$&$\num{1.39}$&$\num{4.2245E-04}$&$\num{0.32}$\\
    &${128}\times{128}$&$\num{9.2047E-04}$&$\num{1.69}$&$\num{1.5609E-04}$&$\num{1.44}$\\
    &${256}\times{256}$&$\num{2.7493E-04}$&$\num{1.74}$&$\num{3.8375E-05}$&$\num{2.02}$\\
    &&&&&\\
    CD2&${32}\times{32}$&$\num{7.9730E-03}$&&$\num{1.2133E-17}$&\\
    &${64}\times{64}$&$\num{2.0322E-03}$&$\num{1.97}$&$\num{1.9345E-17}$&\\
    &${128}\times{128}$&$\num{5.1359E-04}$&$\num{1.98}$&$\num{3.7870E-17}$&\\
    &${256}\times{256}$&$\num{1.2875E-04}$&$\num{2.00}$&$\num{7.4208E-17}$&\\
    &&&&&\\
    WENO5-LLF&${32}\times{32}$&$\num{4.3520E-03}$&&$\num{9.5700E-05}$&\\
    &${64}\times{64}$&$\num{5.2244E-04}$&$\num{3.06}$&$\num{2.2263E-05}$&$\num{2.10}$\\
    &${128}\times{128}$&$\num{3.0996E-05}$&$\num{4.08}$&$\num{8.8323E-07}$&$\num{4.66}$\\
    &${256}\times{256}$&$\num{1.5011E-06}$&$\num{4.37}$&$\num{3.4229E-08}$&$\num{4.69}$\\
    &&&&&\\
    WENO5-Roe&${32}\times{32}$&$\num{2.9596E-03}$&&$\num{8.9004E-05}$&\\
    &${64}\times{64}$&$\num{2.6993E-04}$&$\num{3.45}$&$\num{1.7274E-05}$&$\num{2.37}$\\
    &${128}\times{128}$&$\num{1.5427E-05}$&$\num{4.13}$&$\num{6.7352E-07}$&$\num{4.68}$\\
    &${256}\times{256}$&$\num{6.8663E-07}$&$\num{4.49}$&$\num{2.6719E-08}$&$\num{4.66}$\\
    &&&&&\\
    CD6&${32}\times{32}$&$\num{6.3801E-04}$&&$\num{1.1525E-17}$&\\
    &${64}\times{64}$&$\num{1.0654E-05}$&$\num{5.90}$&$\num{2.4753E-17}$&\\
    &${128}\times{128}$&$\num{1.7796E-07}$&$\num{5.90}$&$\num{5.8132E-17}$&\\
    &${256}\times{256}$&$\num{4.9065E-09}$&$\num{5.18}$&$\num{2.0371E-16}$&\\
    \hline
  \end{tabular}
\end{table}

\tgreen{
We analyze the spatial accuracy of \torange{the} four proposed schemes (JST-LLF, JST-Roe, WENO5-LLF, and WENO5-Roe) using the magnetized iso-density vortex problem proposed by Balsara \cite{2004ApJS..151..149B}.
As argued in \cite{2008JCoPh.227.8209D}, the Gaussian taper of variables in the original problem produces a small error near the boundary, which may reduce the convergence rate of higher-order schemes.
To avoid this effect, we chose a version of this problem described in Mignone et al. \cite{2010JCoPh.229.5896M}.
The initial condition is described by $\rho=1$, $(V_1,V_2)=(1,1)+(-x_2,x_1){\kappa}\mathrm{e}^{q(1-r^2)}$, $(B_1,B_2)=(-x_2,x_1){\mu}\mathrm{e}^{q(1-r^2)}$, and $p=1+[\mu^2(1-qr^2)-\kappa^2\rho]\mathrm{e}^{2q(1-r^2)}/(4q)$, where $r=\sqrt{x_1^2+x_2^2}$, $\kappa=1/(2\pi)$, and $\mu=1/(2\pi)$.
In the actual implementation, we use the magnetic vector potential $A_3={\mu}\mathrm{e}^{q(1-r^2)}/(2q)$ to compute the magnetic field as $(B_1,B_2)=(\delta{A_3}/\delta{x_2},-\delta{A_3}/\delta{x_1})$ so that the discretization error of $\divB$ becomes zero in the initial condition.
The simulation evolves over a time of $10$ units in the two-dimensional domain of $[-5,5]\times[-5,5]$ with periodic boundary conditions.
Following \cite{2010JCoPh.229.5896M}, we set $q=0.5$ for the second-order schemes (JST-LLF and JST-Roe) and $q=1.0$ for the fifth-order schemes (WENO5-LLF and WENO5-Roe).
We used the Courant number $\sigma_\mathrm{CFL}=0.3$ in all cases.
}

\tgreen{
Table \ref{table:vortex} summarizes the $L_1$ errors and corresponding convergence rates measured in $B_1$ and $\divB$.
The $L_1$ error of a quantity $U$ is computed as an average of the absolute value of the difference between the initial and final solutions over the whole domain.
All schemes converge to the designed order of accuracy in both $B_1$ and $\divB$.
The errors and convergence rates of the magnetic field $B_1$ can be directly compared to Table 3 of \cite{2010JCoPh.229.5896M}.
As expected, our second-order schemes provide larger errors than the third-order schemes described in \cite{2010JCoPh.229.5896M}.
The errors and convergence rates by the second-order total variation diminishing scheme in Table 5 of \cite{2004ApJS..151..149B} seems to be comparable or slightly worse than our JST-LLF scheme.
Our fifth-order schemes also provides larger $L_1$ errors than the fifth-order schemes in \cite{2010JCoPh.229.5896M}.
This difference is caused by the larger numerical diffusion by the enhancement of the filtering flux (by setting $c_\mathrm{diff}=2$) and the less diffusive reconstruction by WENO-Z \cite{2008JCoPh.227.3191B} or MP5 \cite{1997JCoPh.136...83S} schemes used in \cite{2010JCoPh.229.5896M}.
}

\tgreen{
For comparison, we also show the results of two additional central difference schemes CD2 and CD6 in Table \ref{table:vortex}.
The CD2 and CD6 schemes are the second- and sixth-order central difference schemes derived by setting the filtering flux to be zero in the JST and WENO5 schemes.
The convergence rates of these central schemes are independent from the filtering scheme and the numerical divergence of the magnetic field.
Both schemes are less diffusive and rapidly converge to the designed order of accuracy.
As the amount of $\divB$ produced by the central difference schemes (CD2 and CD6) is limited by the machine epsilon of the floating-point operation, the order of convergence for $\divB$ is not shown.
The slight reduction in the convergence rate between $128\times128$ and $256\times256$ grid points in the CD6 scheme is caused by the third-order error of the time integration scheme.
If we compute the case of $256\times256$ grid points of the CD6 scheme with smaller CFL number of $0.075$, the $L_1$ error and convergence rate of $B_1$ is $\num{2.7752E-09}$ and $6.00$, respectively.
}


\subsection{One-dimensional hydrodynamic shock tube problems}

\begin{figure}[!ht]
  \centering
  \plotone{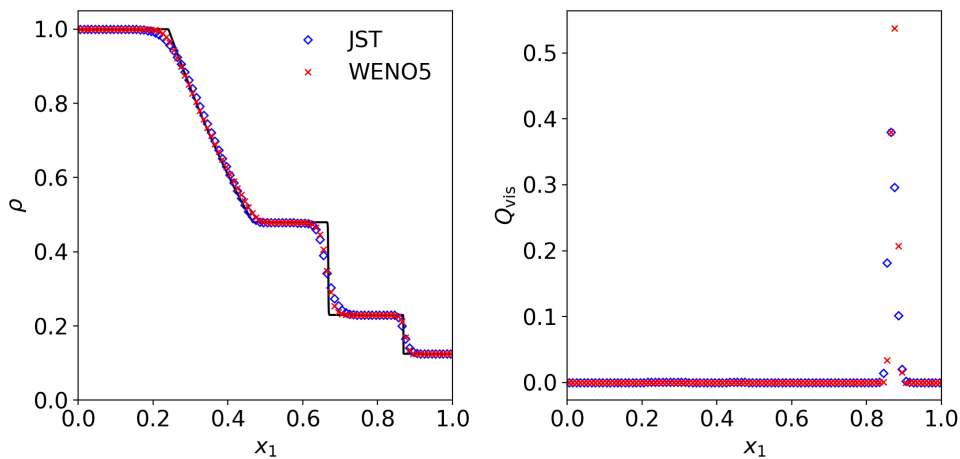}
  \caption{
  One-dimensional Sod's shock tube problem using 100 \tblue{grid points}.
  Shown are the mass density (left panel)
  and the viscous heating rate (right panel) at $t=0.2$
  for \torange{the} JST-Roe scheme (blue diamond)
  and \torange{the} WENO5-Roe scheme (red cross).
  The solid lines indicate the reference solution
  \torange{calculated with the} WENO5-Roe scheme using 4000 \tblue{grid points}.
  \label{fig:sod_shape}
  }
\end{figure}

\torange{
  The hydrodynamic Sod's shock tube problem \cite{1978JCoPh..27....1S} is used to measure the performance of the proposed schemes near the shock wave and discontinuity.
}
Figure \ref{fig:sod_shape} shows the one-dimensional simulation of the Sod's problem calculated with $100$ \tblue{grid points}.
The proposed schemes (JST-Roe and WENO5-Roe) successfully captured the shock front, contact discontinuity, and rarefaction wave.
One advantage of the proposed scheme is that the numerical viscous heating can be derived explicitly.
The result indicates that the viscous heating was concentrated near the shock front near $x{\sim}0.9$, with \torange{a} positive sign indicating the heating process.
The viscous heating was negligible at the contact discontinuity or rarefaction wave.

\begin{figure}[!ht]
  \centering
  \plotone{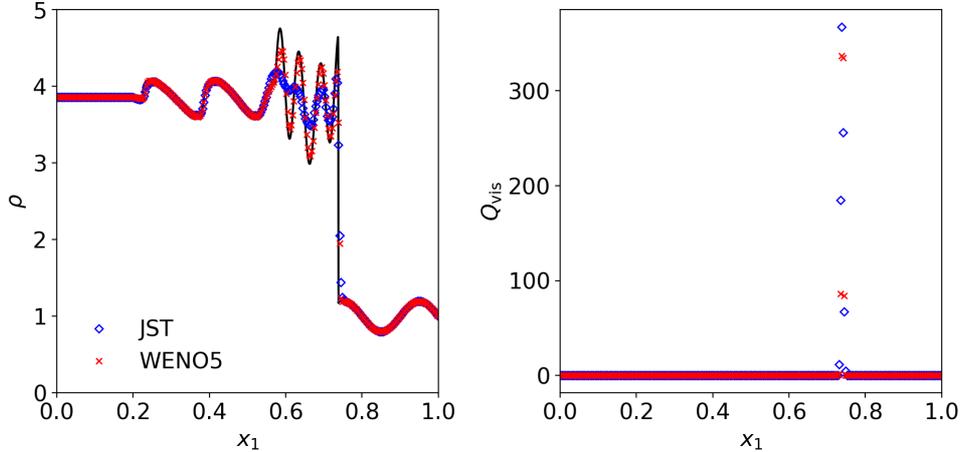}
  \caption{
  One-dimensional Shu--Osher's shock tube problem using 300 \tblue{grid points}.
  Shown are the mass density (left panel)
  and the viscous heating rate (right panel) at $t=0.18$
  for \torange{the} JST-Roe scheme (blue diamond)
  and \torange{the} WENO5-Roe scheme (red cross).
  The solid lines indicate the reference solution
  \torange{calculated with the WENO5-Roe} scheme using 4000 \tblue{grid points}.
  \label{fig:shu_shape}
  }
\end{figure}

The advantage of the higher-order scheme can be apparent in the Shu--Osher's shock tube problem \cite{1989JCoPh..83...32S}.
Figure \ref{fig:shu_shape} shows the results of the one-dimensional Shu-Osher's test using 300 \tblue{grid points}.
The higher-order WENO5-Roe scheme could resolve the wavy pattern after the shock front better than the JST-Roe scheme.
The viscous heating is concentrated near the shock front at $x{\sim}0.7$.

\subsection{One-dimensional MHD shock tube problems}

In the one-dimensional MHD test problems, the discretization error of the magnetic field divergence becomes zero.
Thus, we can assess the performance of the proposed schemes without the influence of the $\divB$ error.

\begin{figure}[!ht]
  \centering
  \plotone{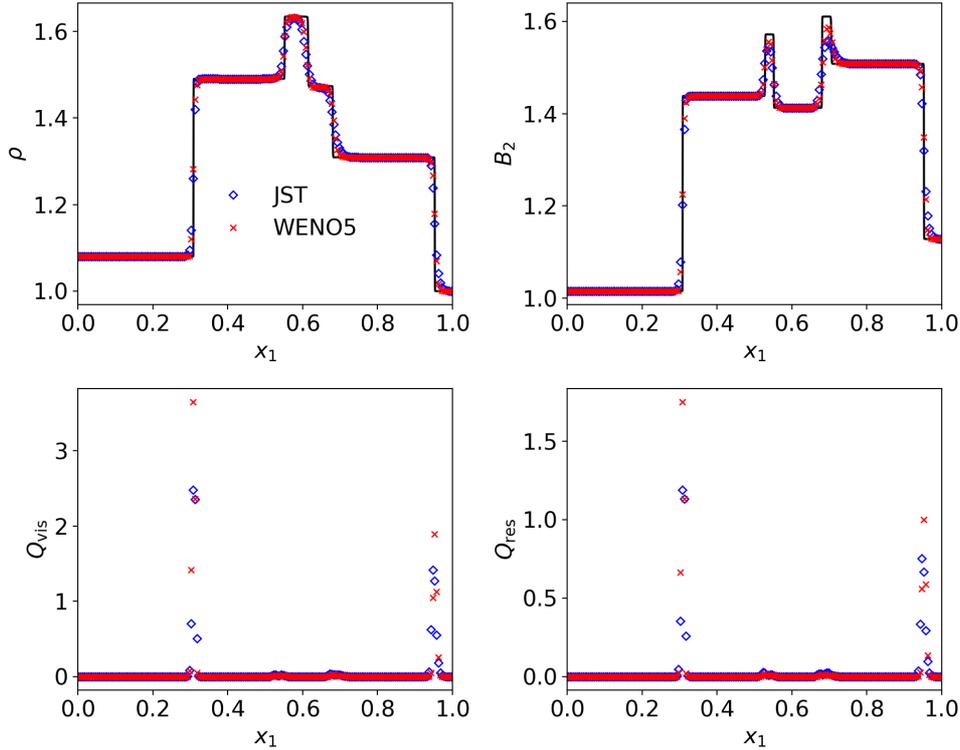}
  \caption{
  One-dimensional Dai--Woodward's shock tube problem using 200 \tblue{grid points}.
  Shown are the mass density (top left panel),
  the y-component of magnetic field (top right panel),
  the viscous heating rate (bottom left panel),
  and the resistive heating rate (bottom right panel) at $t=0.2$
  for \torange{the} JST-Roe scheme (blue diamond)
  and \torange{the} WENO5-Roe scheme (red cross).
  The solid lines indicate the reference solution
  \torange{calculated with the} WENO5 scheme using 4000 \tblue{grid points}.
  \label{fig:dw_shape}
  }
\end{figure}

The Dai--Woodward's MHD shock tube problem \cite{1998ApJ...494..317D} involves various shock waves and discontinuities (i.e., the fast shocks, rotational discontinuities, and slow shocks propagating from each side of the contact discontinuity).
Figure \ref{fig:dw_shape} shows \torange{the} results \torange{of} the one-dimensional Dai-Woodward shock tube problem with 200 \tblue{grid points}.
The JST-Roe and WENO5-Roe schemes were sufficiently robust and captured all shocks and discontinuities in this problem.
No visible numerical overshoots were observed.
The viscous and resistive heating rates are prominently located near the two fast shock fronts.

\begin{figure}[!ht]
  \centering
  \plotone{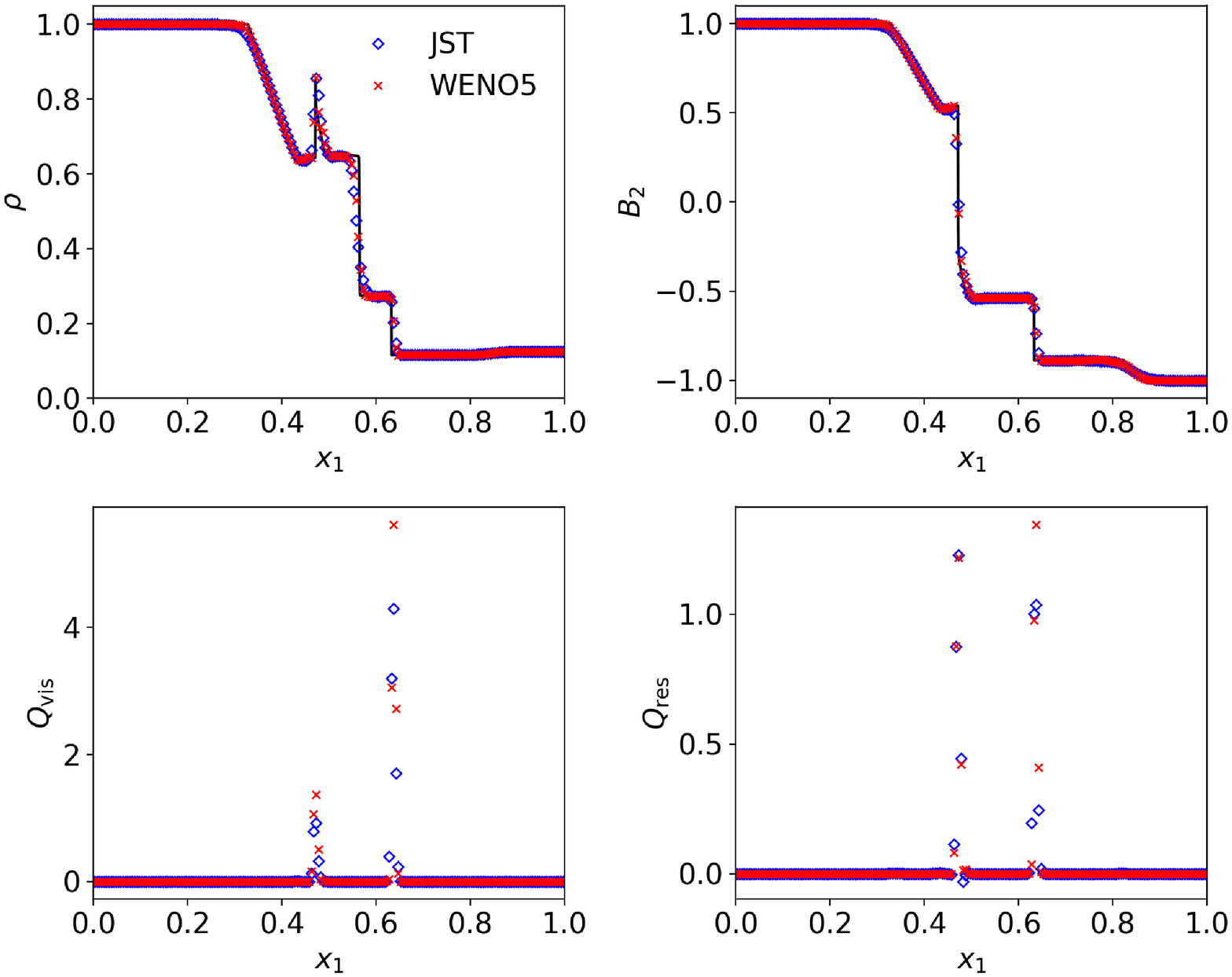}
  \caption{
  One-dimensional Brio--Wu's shock tube problem using 200 \tblue{grid points}.
  Shown are the mass density (top left panel),
  the y-component of magnetic field (top right panel),
  the viscous heating rate (bottom left panel),
  and the resistive heating rate (bottom right panel) at $t=0.1$
  for \torange{the} JST-Roe scheme (blue diamond)
  and \torange{the} WENO5-Roe scheme (red cross).
  The solid lines indicate the reference solution
  \torange{calculated with the} WENO5 scheme using 4000 \tblue{grid points}.
  \label{fig:bw_shape}
  }
\end{figure}

The second MHD shock tube problem considered in this study is the MHD analog of the Sod's shock tube problem originally introduced by Brio \& Wu \cite{1988JCoPh..75..400B}, with the specific heat ratio $\gamma=2$.
Our test problem is a variant of this problem with $\gamma=5/3$ presented by Ryu \& Jones \cite{1995ApJ...442..228R}.
The numerical solution obtained with 200 \tblue{grid points} is shown in Figure \ref{fig:bw_shape}.
This problem involves so-called slow compound shock located near $x{\sim}0.5$.
The numerical solutions calculated with both the JST-Roe and WENO5-Roe schemes agree well with the previously reported results.
Most of the viscous and resistive heating rates appeared near the compound shock and slow shock at $x{\sim}0.65$.
Note that the ratio between the viscous and resistive heating rates depends on the dissipation methods (e.g., LLF or Roe).

\subsection{Two-dimensional MHD Orszag-Tang vortex problem}

\begin{figure}[!ht]
  \centering
  \plotone{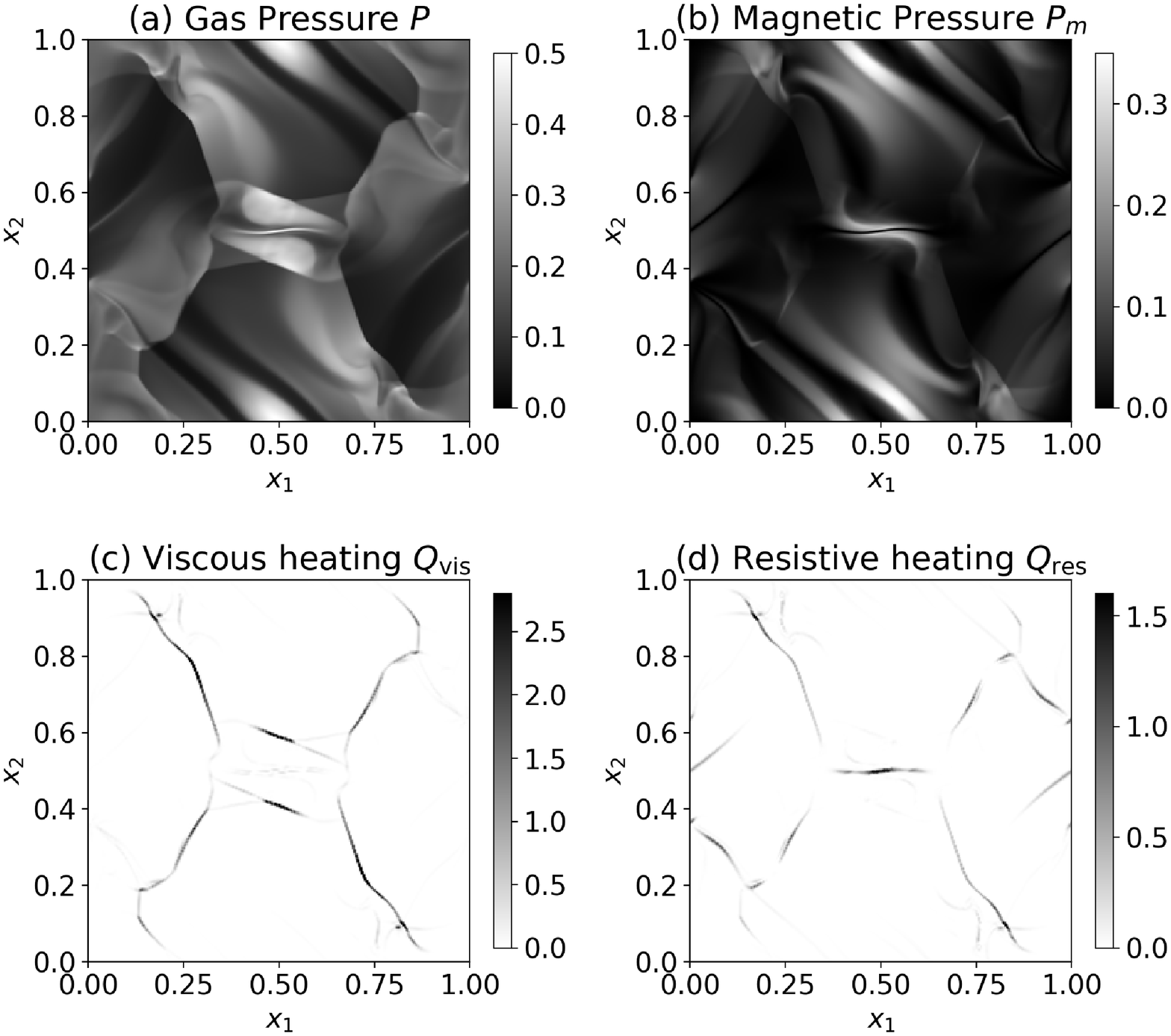}
  \caption{
  Two-dimensional Orszag--Tang vortex problem
  calculated with \torange{the} WENO5-Roe scheme using 256$\times$256 \tblue{grid points}.
  Shown are (a) the gas pressure,
  (b) the magnetic pressure,
  (c) the viscous heating rate,
  and (d) the resistive heating rate at $t=0.48$.
  \label{fig:ot_shape}
  }
\end{figure}

Multi-dimensional MHD problems \tblue{suffer} from the numerical error in the divergence of the magnetic field.
The robustness and energy-conservation of the proposed schemes in the two-dimensional domain are tested \torange{using} the Orszag--Tang vortex problem \cite{1979JFM....90..129O}.
There are minor variations of the numerical settings in the previous literatures.
The numerical setup used in this paper is identical to that of Ryu et al. \cite{1995ApJ...452..785R,1998ApJ...509..244R}.
The numerical solution is shown in Fig. \ref{fig:ot_shape}.
The spatial profiles of gas and magnetic pressures agreed well with the previous reports (e.g., Fig. 3 of Ryu et al. \cite{1998ApJ...509..244R}).
The viscous and resistive heating rates were concentrated near the shock fronts.

\begin{figure}[!ht]
  \centering
  \plotsml{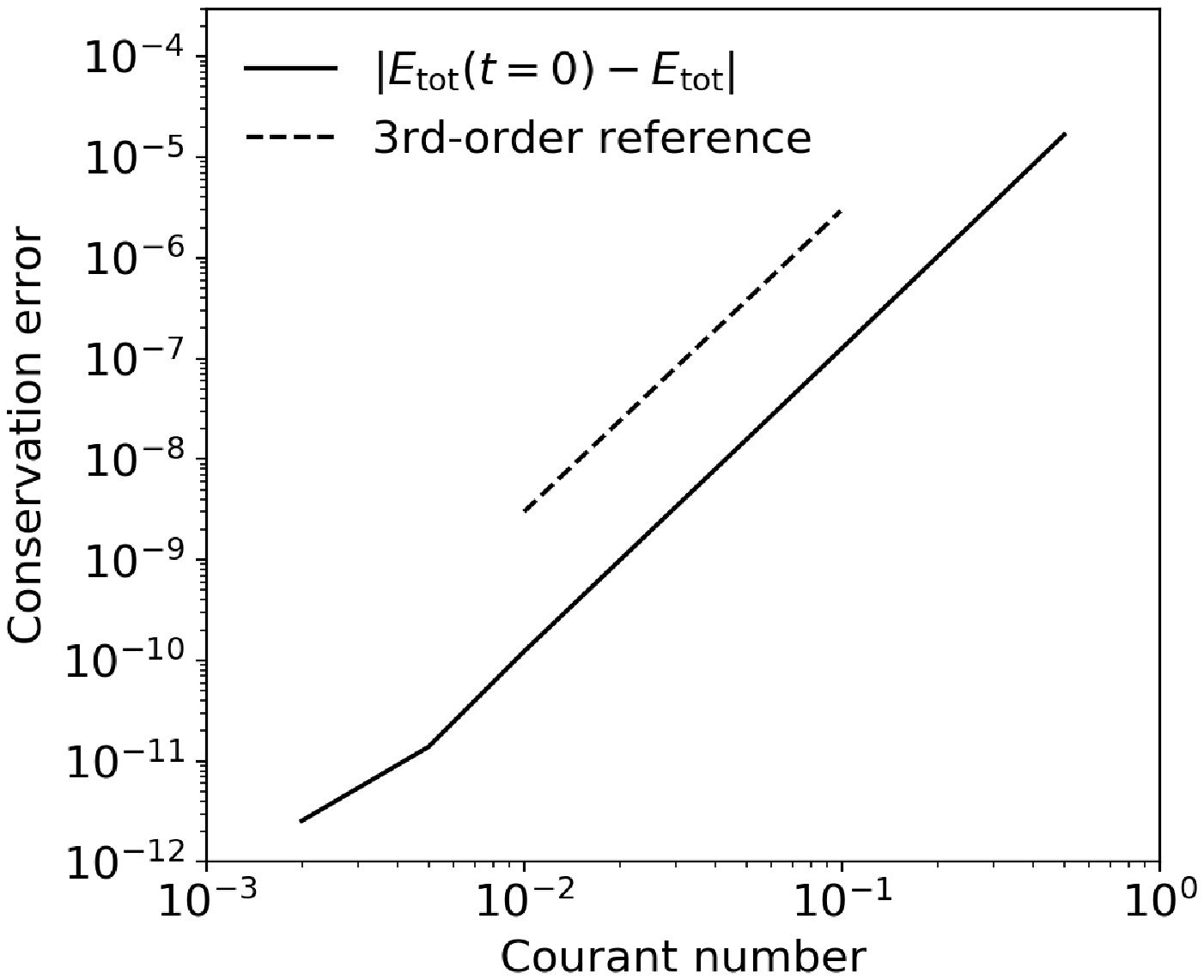}
  \caption{
  Conservation error of the total energy for Orszag--Tang vortex problem
  calculated with \torange{the} WENO5-Roe scheme using 128$\times$128 \tblue{grid points}.
  Shown is difference between the initial ($t=0$) and final step ($t=0.48$)
  of the volume-averaged total energy density.
  \label{fig:ot_cnvrg}
  }
\end{figure}

When the total energy equation is not directly solved in the MHD schemes, the total energy conservation may be numerically violated.
The proposed scheme is designed to satisfy the total energy conservation in terms of the spatial discretization.
However, the total energy conservation may be violated by the numerical error of the temporal discretization \torange{as discussed in Sec. \ref{subsec:advection}}.
\torange{
In our formulation, the conservation error of the total energy depends only on the discretization error of the time integration scheme.
Because the third-order Runge-Kutta method was used for time integration, the conservation error of the total energy should converge by third order against the size of time step used in the simulations.
}
The convergence of the total energy conservation in terms of the Courant number of the time step criterion is shown in Fig. \ref{fig:ot_cnvrg}.
Here, we employed the lower spatial resolution with $128\times128$ \tblue{grid points} to reduce the computational cost. 
The observed third-order convergence of the conservation error of the total energy is consistent \torange{with our expectation}.

We also checked the conservation error of the momentum.
\torange{Due to} the spatial symmetry of this test problem, the conservation error of the momentum was kept constant in the \torange{round-off error}.
Instead, we evaluated the physical part of the Lorentz force Eq. \ref{eq:anl_fl_ncnsv} and the artificial field-aligned force $B_i\divB$.
The standard deviation (in time and space) of the $x$-component of the field-aligned force was several tens of \torange{percent} of the physical part of the Lorentz force.
The amount of this field-aligned force is a direct consequence of the numerical $\divB$ error, which depends on the filtering scheme, the spatial resolution, and the parameters of the divergence cleaning \torange{method} ($c_\psi$ and $\tau_\psi$ in this study).
As we discussed in \torange{Sec.} \ref{subsec:divb}, the present scheme can be modified so that the momentum conservation is satisfied even when $\divB$ is nonzero.
This alternative version may be used in the problems where the strict momentum conservation by the Lorentz force is required.


\begin{figure}[!ht]
  \centering
  \plotbig{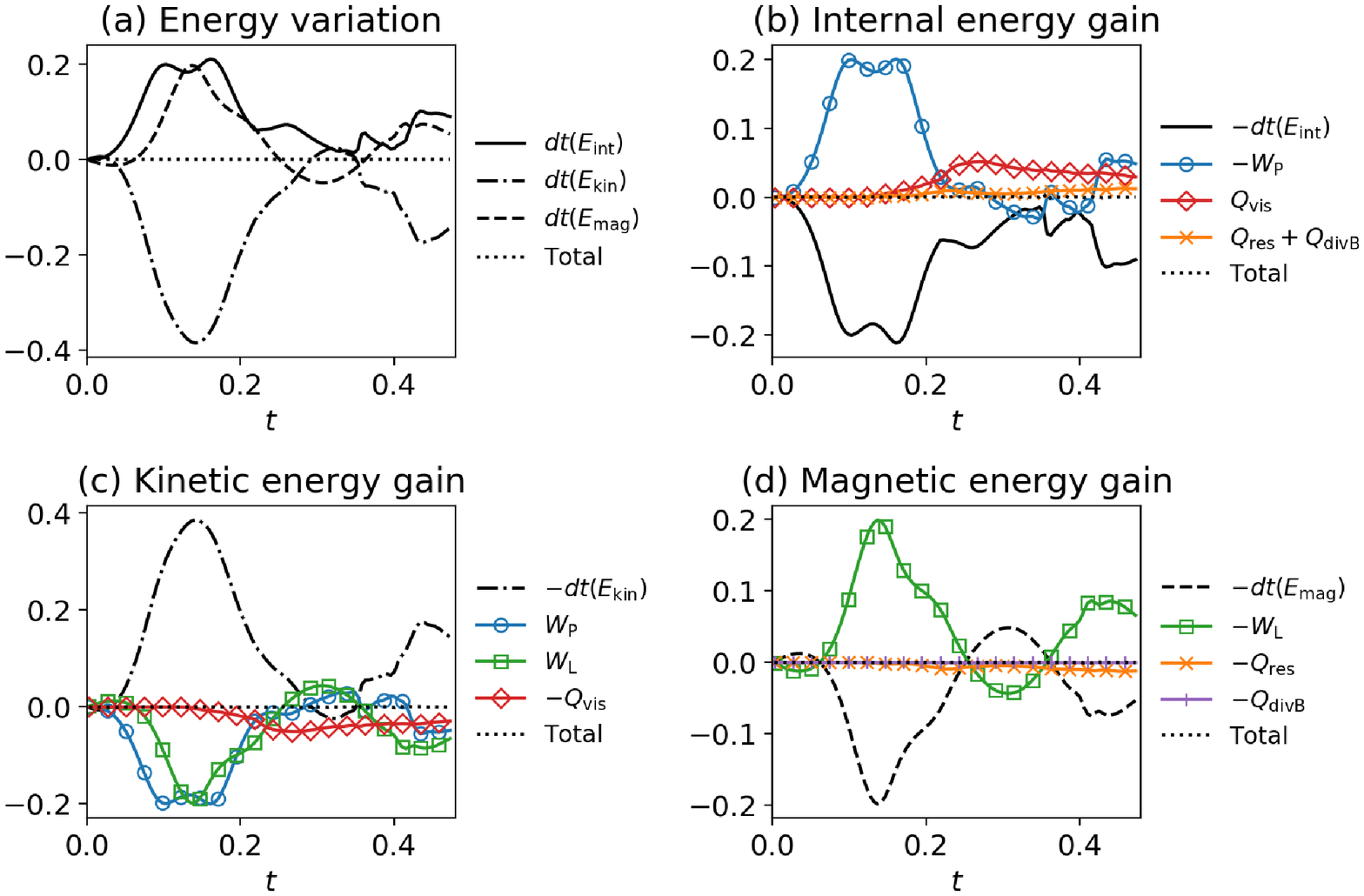}
  \caption{
  Contributions in energy equations for Orszag--Tang vortex problem
  calculated with \torange{the} WENO5-Roe scheme using 256$\times$256 \tblue{grid points}.
  Shown are
  (a) the temporal variation of volume-averaged energy densities,
  (b) the gain and loss of internal energy,
  (c) the gain and loss of kinetic energy,
  and (d) the gain and loss of magnetic energy.
  \label{fig:ot_heat}
  }
\end{figure}

One of the advantages of the proposed scheme is that the analysis of the energy variations in the numerical solution can be accurate and straightforward.
Figure \ref{fig:ot_heat} shows the temporal variation of the internal, kinetic, and magnetic energies and the contributions to \torange{their} variation.
The initial kinetic energy is converted into the internal and magnetic energies during the time evolution.
The prominent energy exchange is through the works by the pressure gradient and Lorentz forces.
\torange{Both the} pressure gradient force \torange{the viscous heating convert} the kinetic energy into the internal energy.
The temporal change of the magnetic energy is dominated by the work done by the Lorentz force.
The resistive heating has only minor effect in the energy conversion.
The heating from the $\divB$ error, i.e., $Q_\psi$, is negligible (less than \torange{$4\times10^{-4}$} as the volume-averaged value).
This small contribution of $Q_\psi$ can be explained that this value is roughly proportional to the square of the divergence of the magnetic field.
For each energy \tblue{equation}, \tblue{all the} contributions are fully balanced, indicating the accuracy of the energy analysis presented here.

\subsection{Two-dimensional MHD blast problem}
\label{subsec:blast}

\begin{figure}[!ht]
  \centering
  \plotone{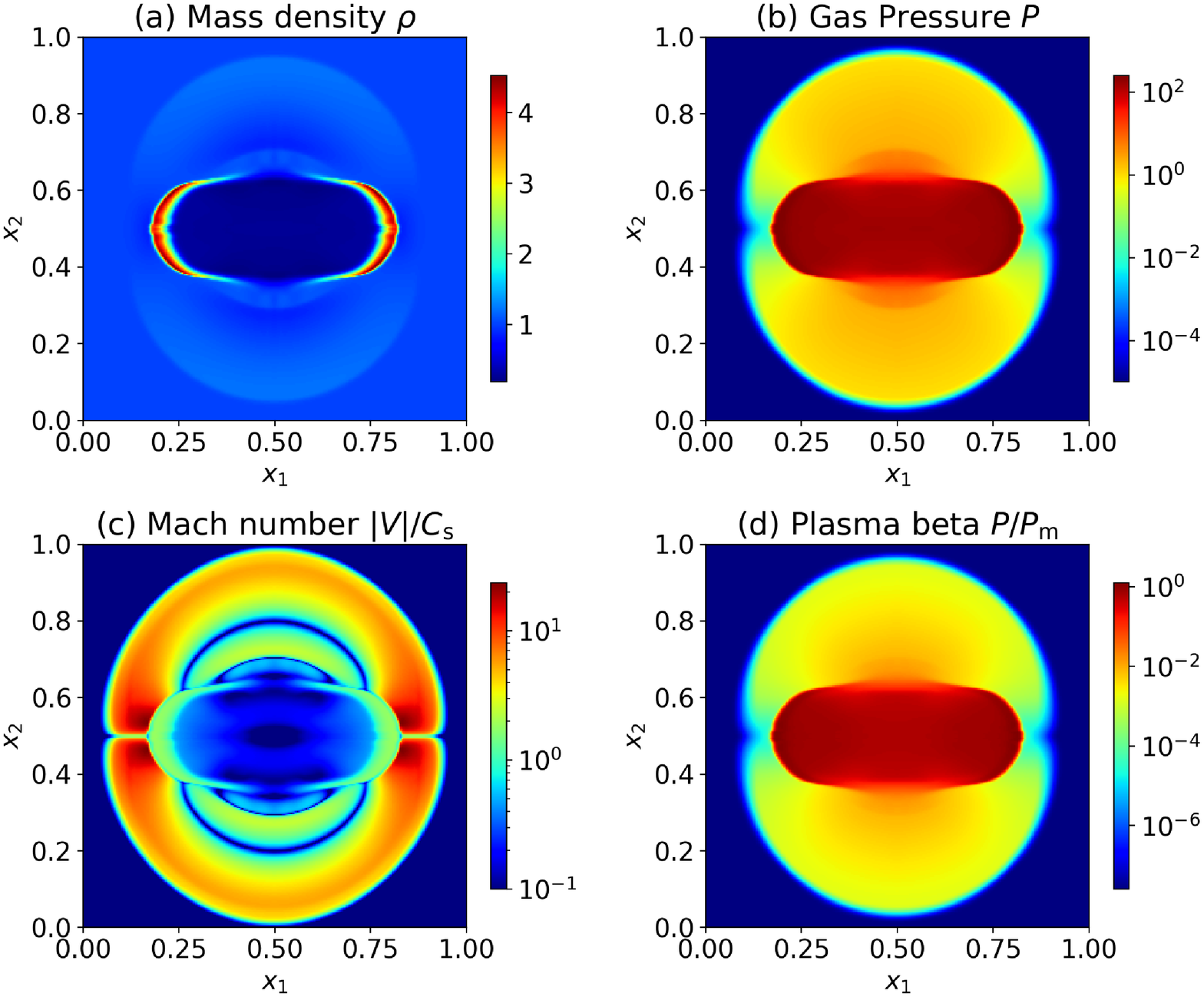}
  \caption{
  Two-dimensional MHD \torange{blast} problem
  calculated with \torange{the} JST-Roe scheme using 256$\times$256 \tblue{grid points}.
  Shown are (a) the mass density,
  (b) the gas pressure,
  (c) the Mach number,
  and (d) the plasma beta at $t=0.01$.
  The range of the color map is determined by the maximum and minimum values at the snapshot (except the lower range of $0.1$ in panel c).
  \label{fig:br_shape}
  }
\end{figure}

The two-dimensional MHD blast wave problem proposed by Balsara \& Spicer \cite{1999JCoPh.149..270B} is one of the most stringent among the standard test problems.
In the original version, the plasma beta (i.e., the ratio between the gas pressure and magnetic pressure) in the ambient is set to approximately $2.5\times10^{-4}$.
\torange{
This problem is a severe numerical benchmark for the MHD solvers because the positivity of the gas pressure can be easily violated due to the shock propagation in the extremely strong magnetic field.
}
We found that the JST-Roe, JST-LLF, and WENO5-LLF schemes can handle this \torange{problem} successfully, \torange{whereas} the WENO5-Roe scheme failed due to the negative pressure during the time integration.

To assess the further robustness of the proposed schemes, we calculated the MHD blast problem in the non-standard setting with a lower plasma beta.
The initially uniform mass density, velocity, and magnetic field are imposed as \torange{$(\rho,V_1,V_2,B_1,B_2)=(1,0,0,100/\sqrt{4\pi},0)$}.
The initial gas pressure is set to $10^{-5}$ except in a central circle of radius $0.1$, where the pressure is set to $1000$.
The only difference from the original problem \cite{1999JCoPh.149..270B} is the much lower gas pressure in the ambient region.
The resultant plasma beta in the ambient region is $2.5\times10^{-8}$, which is the most stringent setup among the family of the MHD blast wave \torange{problems} \tred{
  tested in previous literatures \cite{1999JCoPh.149..270B,2008ApJS..178..137S,2009JCoPh.228.5040B,2013JCoPh.243..269L}.
}
The numerical solution obtained by $256\times256$ \tblue{grid points} with JST-Roe scheme is shown in Fig. \ref{fig:br_shape}.
The JST-Roe and JST-LLF schemes can handle this problem, but WENO5-LLF scheme causes a negative pressure during the time integration.
We found that \torange{the} two JST schemes are extremely robust in this problem and stable even when we set the ambient gas pressure to \torange{$10^{-7}$} (i.e., the ambient plasma beta of \torange{$2.5\times10^{-10}$}).

\subsection{Two-dimensional MHD Rotor problem}
\label{subsec:rotor}

\begin{figure}[!ht]
  \centering
  \plotone{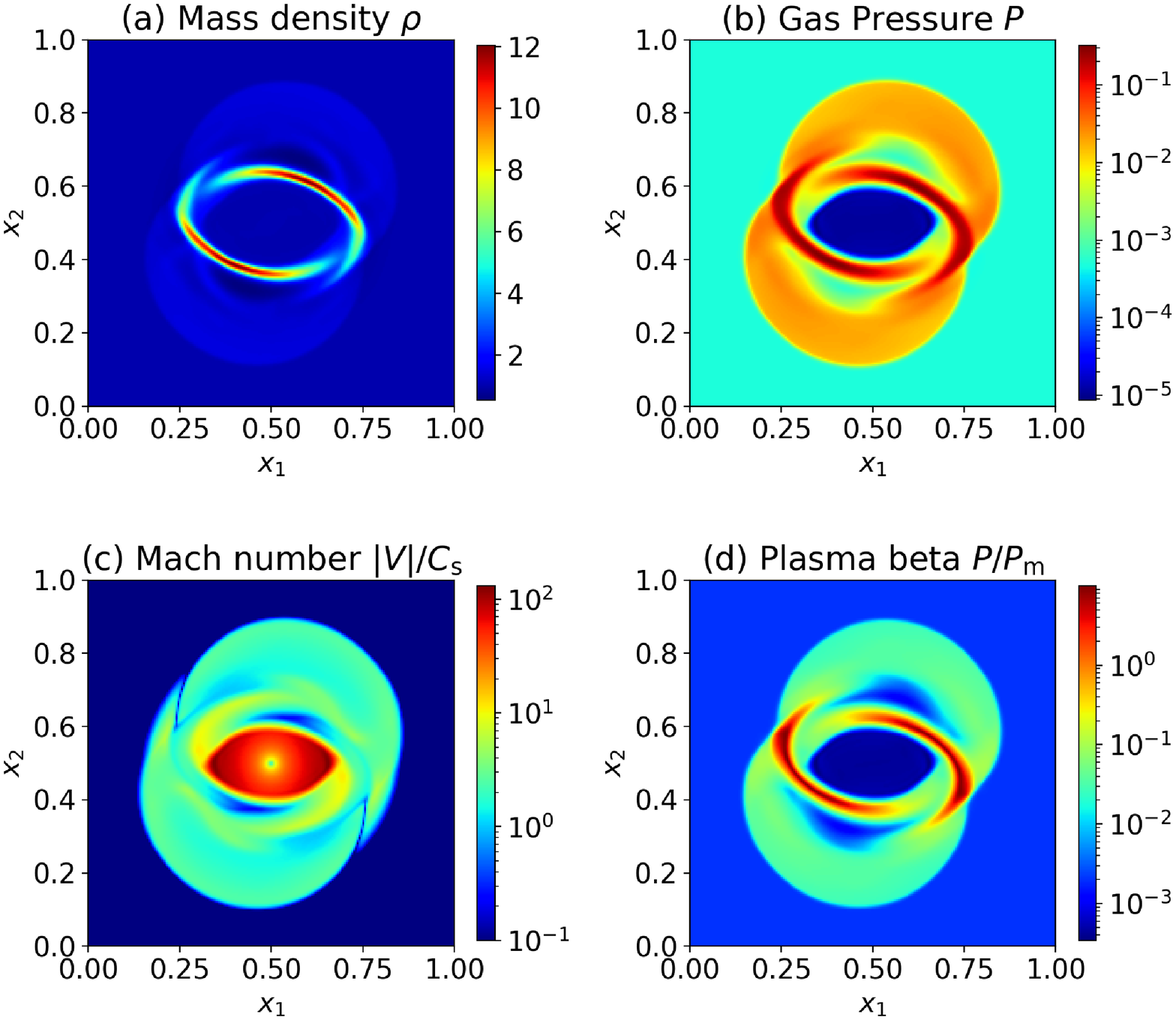}
  \caption{
  Two-dimensional MHD rotor problem calculated with \torange{the} WENO5-LLF scheme using 200$\times$200 \tblue{grid points}.
  Shown are (a) the mass density, (b) the gas pressure, (c) the Mach number,
  and (d) the plasma beta at $t=0.295$.
  The range of the color map is determined by the maximum and minimum values at the snapshot (except the lower range of $0.1$ in panel c).
  \label{fig:rt_shape}
  }
\end{figure}

The two-dimensional MHD rotor problem originally proposed by Balsara \& Spicer \cite{1999JCoPh.149..270B} involves a rotating disk with an initial magnetic field perpendicular to the rotation axis.
The strong rotational discontinuities along with the shocks and rare factions are generated by the shearing and expansion motion of the rotating disk.
Among the several variations of this problem, the second rotor test described by T{\'o}th \cite{2000JCoPh.161..605T} was selected for this study.
All of the proposed schemes (JST-Roe, JST-LLF, WENO5-Roe, and WENO5-LLF) are capable of solving this problem.

\torange{A} non-standard alternative of the rotor problem with more stringent parameter \torange{was also carried out}.
The numerical setup is identical to that of the T{\'o}th's second rotor problem \cite{2000JCoPh.161..605T} except the initial gas pressure and linear taper.
The initial gas pressure is uniformly set to $5\times10^{-4}$ whereas it is $0.5$ in the original problem.
The initial plasma beta is about $2.0\times10^{-3}$.
The linear taper is applied between a radius of $0.1$ and a radius of $0.13$ so that the rotor's velocity and mass density linearly approach \torange{to the background} over six \tblue{grid points} when the problem is solved with $200\times200$ \tblue{grid points}.
Figure \ref{fig:rt_shape} shows the numerical solution solved by \torange{the} WENO5-LLF scheme.
The expansion of the rotating disk caused \torange{the region with very low pressure} near the center of the numerical domain.
At the final snapshot ($t=0.295$), the plasma beta was lower than $2\times10^{-4}$ and the local Mach number exceeds $150$ in the \torange{low-pressure} region.
\tred{
  These values of the Mach number and plasma beta are more severe than those used in previous studies \cite{1999JCoPh.149..270B,2008ApJS..178..137S,2009JCoPh.228.5040B,2013JCoPh.243..269L}, making the problem a severe benchmark for the numerical robustness.
}

We found that two JST schemes and the WENO5-Roe scheme failed \torange{to handle} this problem with the initial gas pressure of $5\times10^{-4}$.
By changing the initial gas pressure from $0.5$ to $5\times10^{-4}$ multiplying a factor of $0.1$ in each step, we checked the minimum gas pressure that the numerical scheme can handle this problem.
The minimum values of the initial gas pressure that can be handled by the JST-LLF, JST-Roe, and WENO5-Roe schemes were $10^{-3}$, $10^{-2}$, and $10^{-1}$, respectively.
The poor robustness of the JST-Roe and WENO5-Roe schemes can be explained that the viscous and resistive heating can be negative for the \torange{Roe-type} model.
The higher robustness of the WENO5-LLF scheme over the JST-LLF scheme may have been caused by the higher gradient of the solution (see panel a in Fig. \ref{fig:rt_shape}).
We found that the \torange{JST-LLF and JST-Roe schemes} can solve the problem with the initial gas pressure of $5\times10^{-4}$ \torange{and $5\times10^{-3}$, respectively,} when the finer numerical mesh of $400\times400$ \tblue{grid points} is used.
\torange{
We found no improvement of the numerical robustness for the two WENO5 schemes with $400\times400$ grid points.
}

\section{Conclusions}
\label{sec:cncl}

We have presented an energy-consistent formulation of the compressible magnetohydrodynamic equations.
The transport and interaction of the internal, kinetic, and magnetic energies are satisfied in the discrete sense, while maintaining the standard conservation property of the mass, momentum, magnetic flux, and total energy.
\tblue{These} characteristics \torange{have} been accomplished through the application of a simple constructive strategy \torange{of using} the discrete versions of the product rule.
The shock capturing was achieved \torange{by introducing} the nonlinear filtering flux in the discretized equations.
The energy-consistent formulation of the nonlinear filtering flux for both hydrodynamic and MHD equations was also suggested.
The viscous and resistive heating rates \torange{become} positive when the filtering flux satisfies the specific conditions.

\tblue{
The proposed formulations were implemented with the spatially second-order and the sixth-order central difference operators.
The filtering flux is developed based the second-order JST scheme and fifth-order WENO reconstruction with the LLF-type and Roe-type filtering schemes.
All of these schemes are integrated by the third-order SSP Runge-Kutta method.
}
\torange{The combination} of the energy-consistent finite difference formulation and the appropriate filtering with small numerical overshoot was shown to yield the excellent robustness for the most stringent problems.

\tblue{
  In this study, we assume the periodic boundary conditions in all test problems presented in Sec. \ref{sec:test}.
  The conservation property of the (magneto)hydrodynamic variables is affected by the non-periodic boundary conditions in practical problems.
  For the practical implementation of the non-periodic boundary conditions in the finite difference schemes with the secondary conservation property, we refer the reader to the discussion in Morinishi \cite{1998JCoPh.143...90M} and Desjardins \cite{2008JCoPh.227.7125D}.
}

\section*{Acknowledgements}

This work was supported by JSPS KAKENHI grant No. JP19K14756.
A major part of numerical computations was carried out on Cray XC50 at Center for Computational Astrophysics, National Astronomical Observatory of Japan.
This work was supported by the computational joint research program of the Institute for Space-Earth Environmental Research (ISEE), Nagoya University.
Some numerical computations were carried out on the FX100/Flow supercomputer system at the Information Technology Center, Nagoya University.
Analysis of the numerical simulations were carried out in the Center for Integrated Data Science, Institute for Space-Earth Environmental Research, Nagoya University through the joint research program.



\end{document}